\documentclass[twocolumn,floatfix,showpacs,preprintnumbers,amsmath,amssymb]{revtex4-1}
\newcommand{\newwidth}{0.245\textwidth}
\newcommand{\newheight}{0.17\textwidth}
\newcommand{\newwidthII}{0.24\textwidth}
\newcommand{\newheightII}{0.19\textwidth}
\newcommand{\newwidthIII}{0.2\textwidth}
\newcommand{\newheightIII}{0.12\textwidth}

\newcommand{\newwidthIX}{0.2\textwidth}
\newcommand{\newheightIX}{0.125\textwidth}
\usepackage{graphicx}
\usepackage{dcolumn}
\usepackage{bm,verbatim}
\usepackage{sidecap}

\begin{document}

\title{Sharp peaks in the conductance of double quantum dot and quantum dot spin-valve systems 
at high temperatures: A hierarchical quantum master equation approach} 

\author{S.\ Wenderoth}
\author{J.\ B\"atge}
\author{R.\ H\"artle}
\affiliation{Institut f\"ur theoretische Physik, Georg-August-Universit\"at G\"ottingen, D-37077 G\"ottingen, Germany. 
}

\date{\today}

\begin{abstract}
We study sharp peaks in the conductance-voltage characteristics of a double quantum dot and a quantum dot 
spin-valve that are located around zero bias. The peaks share similarities with a Kondo peak but can be clearly 
distinguished, in particular as they occur at high temperatures. The underlying physical mechanism 
is a strong current suppression that is quenched in bias-voltage dependent ways by exchange interactions. 
Our theoretical results are based on the quantum master equation methodology, including the 
Born-Markov approximation and a numerically exact, hierarchical scheme, which we extend here to the spin-valve case. 
The comparison of exact and approximate results allows us to reveal the 
underlying physical mechanisms, the role of first-, second- and beyond-second-order processes and 
the robustness of the effect. 
\end{abstract}

\pacs{85.35.-p, 73.63.-b, 73.40.Gk}

\maketitle

\emph{Introduction:} Exchange interactions are ubiquitous in quantum many-body physics. 
They originate from the symmetry of many-body wave functions with respect to the exchange of 
indistinguishable particles \cite{Heisenberg1926,Dirac1926}. Imprints of exchange interactions can be found 
in the energy spectrum of atoms 
and molecules, scattering cross sections and by the existence of magnetic phases \cite{Goodenough1966,White2007}. 
The description of exchange interactions represents a major challenge. Yet, an exact treatment is crucial. 
This can be seen, \emph{e.g.}, by the huge effort that is spent in density functional theory 
to find the correct exchange-correlation functional \cite{KohnSham1965,EngelDreizler2011}. 
Another prominent example is the Kondo effect \cite{Kondo1964,Hewson93}, 
where the spin of an impurity is screened by itinerant electrons. 
Here, exchange interactions are attractive and lead to the formation of a bound state, the Kondo singlet. 
Its binding energy is referred to as the Kondo temperature $T_{\text{K}}$.

\begin{figure}
\begin{tabular}{lll}
\resizebox{\newwidthIX}{\newheightIX}{
\includegraphics{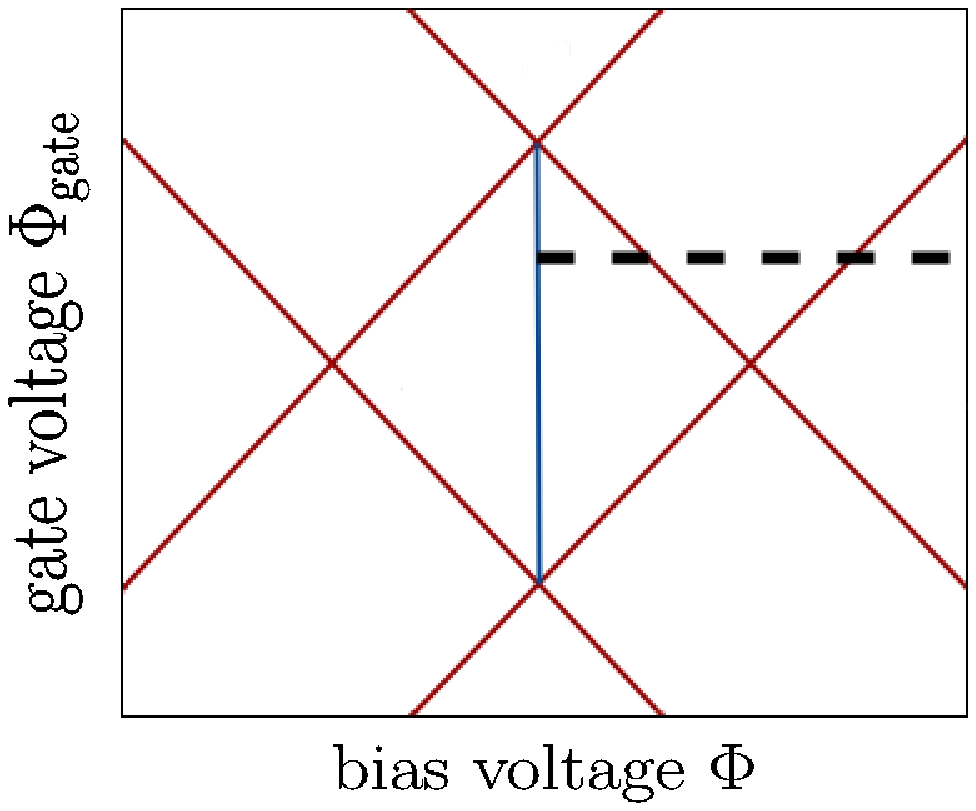} }
&
\hspace{-1.1cm}\resizebox{\newwidthIII}{\newheightIII}{
\includegraphics{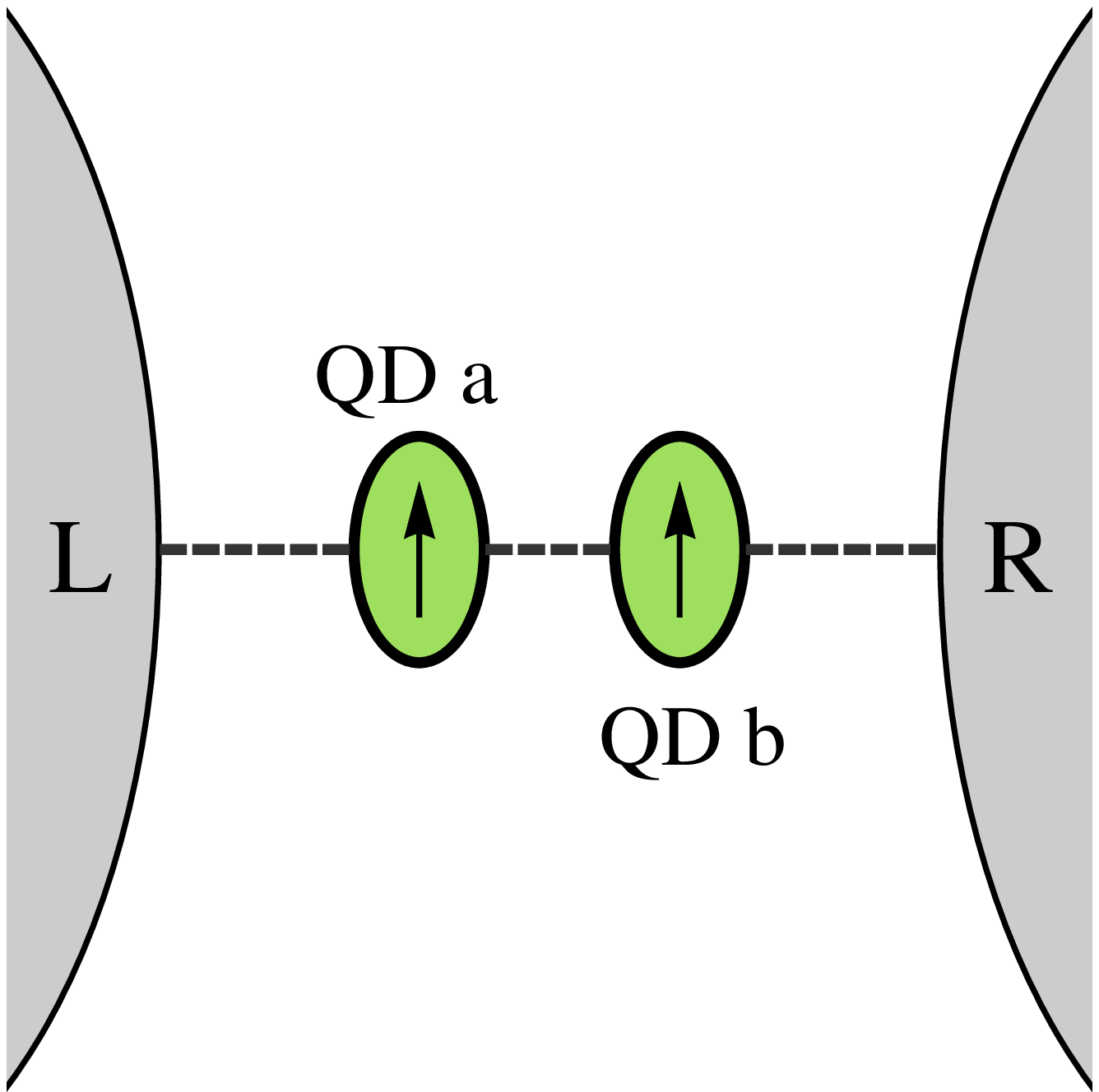} }
&
\hspace{-0.9cm}\resizebox{\newwidthIII}{\newheightIII}{
\includegraphics{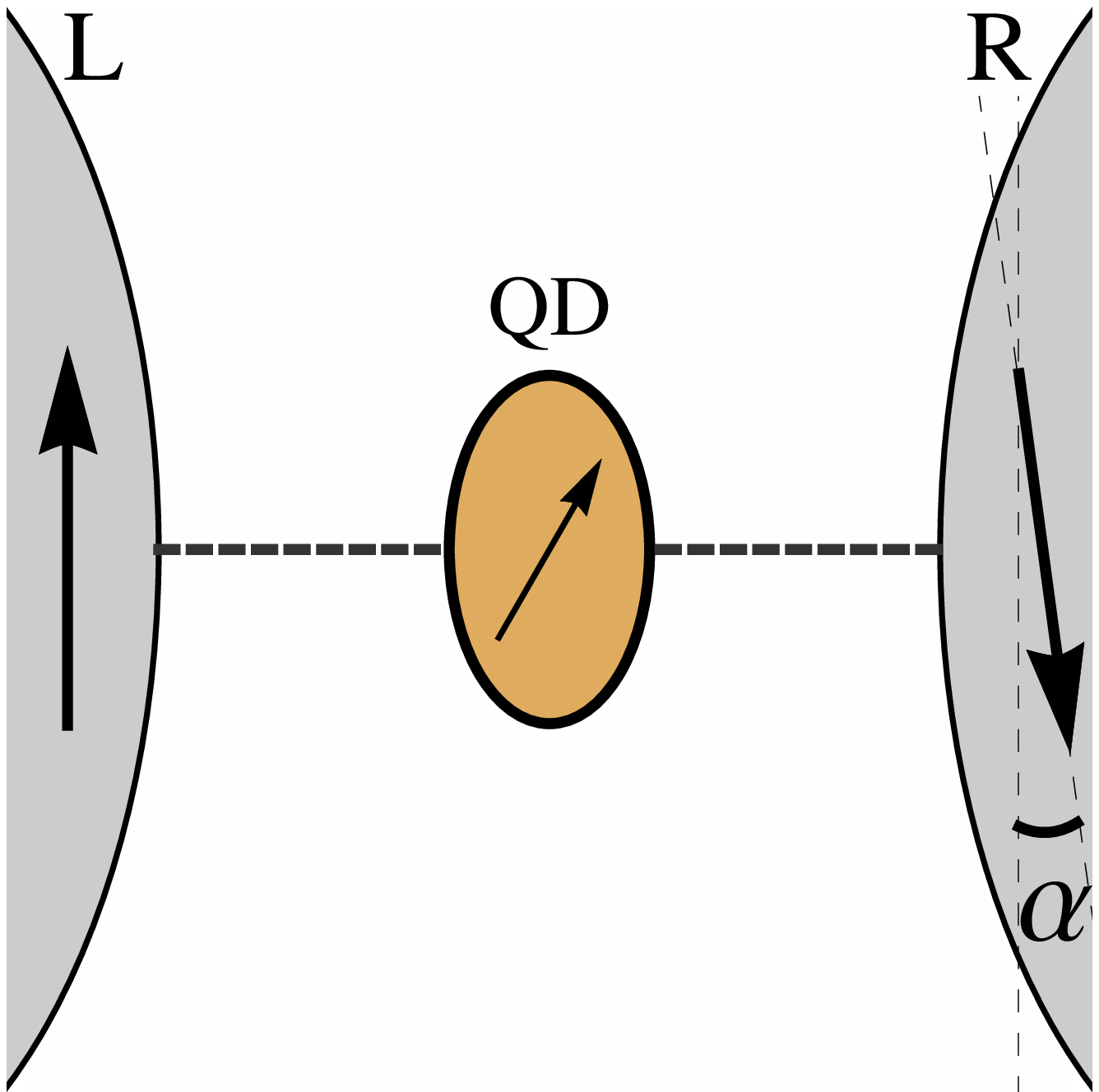}  
}
\\
\end{tabular}
\caption{\label{qdfig} 
Left panel: Scheme of a typical conductance map, where the conductance is plotted as a function 
of both bias and gate voltage. The red diagonals are associated with tunneling through 
the single-particle levels $\epsilon_{m}$ and $\epsilon_m + U$. 
The blue line depicts a zero-bias peak. It occurs for gate voltages $\Phi_{\text{gate}}$ 
where the impurity or quantum dot system is populated by an odd number of electrons on average. 
The dashed black line indicates the position of the conductance-voltage characteristics 
depicted in Fig.\ \ref{effectofhigherordershifted}. 
Middle panel: Double quantum dot (DQD) where dot a is coupled to the left (L) and dot b to the right electrode (R). 
Right panel: 
Quantum dot spin-valve (SV) with non-collinearly polarized electrodes. 
The angle between the polarization vectors is denoted by $\alpha$. 
}
\end{figure}

In normal metals like gold, Kondo singlets 
can lead to an increase of the resistivity at low temperatures \cite{deHaas1934,Kondo1964}. 
But they can also strongly enhance transport such as, for example, in 
single-molecule junctions \cite{Liang02,Pasupathy2004,Natelson2005,Osorio2007,Mugarza2011,Rakhmilevitch2013,Wagner2013} or 
quantum dot devices \cite{Goldhaber1998,Cronenwett1998,Wiel2000,Jeong2001}, 
where they establish a direct link between the electrodes. The associated 
conductivity is the same as the one of a ballistic channel 
and given by the conductance quantum $G_0=2e^2/h\approx7.75\cdot10^{-5}$\,A/V \cite{Wiel2000}. 
The underlying processes need to be coherent such that the effect is found only at low 
temperatures $T\lesssim T_{\text{K}}$ or, equivalently, small bias voltages where the range of electrons 
involved in the singlet formation is narrow enough. This leads to very sharp peaks (Kondo peaks)
in the corresponding conductance-voltage characteristics at zero bias 
(cf.\ the blue line of the schematic conductance map shown in the left panel of Fig.\ \ref{qdfig}).

In this article, we focus on another, complementary transport phenomenon due to exchange interactions. 
It also leads to sharp peaks in the conductance-voltage characteristics at or near zero bias. 
Such peaks were reported only recently for double quantum dot (DQD) structures \cite{Hartle2013b} (see middle panel of Fig.\ \ref{qdfig}) and 
quantum dot spin-valve (SV) setups \cite{Hell2014} (see right panel of \text{Fig.\ \ref{qdfig}}). 
Here, we develop a physically intuitive picture, showing that the peak structures 
can be understood in both systems on the same footing, 
namely due to orbital and spin exchange processes. 
We also demonstrate how the peaks can be distinguished from Kondo peaks by experimentally accessible means, 
in particular their position in asymmetric scenarios \cite{Hell2014} and a non-monotonic temperature dependence with 
a power-law scaling $\sim 1/T^2$ at high temperatures.  
In addition, we substantially extend previous considerations, which were based on 
first and second order perturbation theory, by numerically exact results. 
To this end, we employ the hierarchical quantum master equation technique (HQME) \cite{Jin2008,Hartle2013b} 
and develop it further to the SV case. We also use approximate Born-Markov results to 
highlight the robustness of the effect.

\emph{Modeling:} Both quantum dot systems 
are naturally described by an impurity part, $H^{A}_{\text{Imp}}$, a lead part, $H^{A}_{\text{L+R}}$, 
and an impurity-lead coupling, $H^{A}_{\text{Imp,L+R}}$ ($A\in\{\text{DQD,SV}\}$). 
The DQD structure (middle panel of Fig.\ \ref{qdfig}) is described by a spinless Anderson impurity model 
with $H^{\text{DQD}}_{\text{Imp}} = \sum_{m\in\{\text{a},\text{b}\}} 
 \epsilon_{m} d_{m}^{\dagger}d_{m} + U d_{\text{a}}^{\dagger}d_{\text{a}} d_{\text{b}}^{\dagger}d_{\text{b}} 
 + ( t_{\text{hopp}} d_{\text{a}}^{\dagger}d_{\text{b}} + h.c.)$, 
$H^{\text{DQD}}_{\text{Imp,L+R}} = \sum_{m,K,k\in K} \nu_{K,m} t_{k} c_{k}^{\dagger}d_{m} + h.c.$ and  
$H^{\text{DQD}}_{\text{L+R}} = \sum_{k\in\{\text{L},\text{R}\}} \epsilon_{k} c_{k}^{\dagger}c_{k}$, 
where we assume that a magnetic field is used to drive one of the spin species out of the system \cite{Wiel2002,Keller2014} 
\footnote{Using units where the elementary charge $e=1$, $\hbar=1$ and the Boltzmann 
constant $k_\text{B}=1$.}. It involves two orbitals localized on quantum dots a and b with hopping amplitude 
$t_{\text{hopp}}$. 
Double occupation of the orbitals requires an additional charging energy $U$. 
The coupling matrix elements $\nu_{K,m}$ and $t_k$ describe the tunneling 
of an electron from dot $m$ into lead $K$ at energy $\epsilon_k$. 
The SV system (right panel of Fig.\ \ref{qdfig}) can be modeled by a (spinful) Anderson impurity that is coupled to non-collinearly 
polarized electrodes with 
$ H^{\text{SV}}_{\text{Imp}} = \sum_{m\in\{\uparrow,\downarrow\}} 
 \epsilon_{m} d_{m}^{\dagger}d_{m} + U d_{\uparrow}^{\dagger}d_{\uparrow} d_{\downarrow}^{\dagger}d_{\downarrow}$, 
 $H^{\text{SV}}_{\text{L+R}} = \sum_{k\in\{\text{L},\text{R}\},m} 
 \epsilon_{mk} c_{mk}^{\dagger}c_{mk}$ and $
 H^{\text{SV}}_{\text{Imp,L+R}} = \sum_{m,m',K,k\in K} \nu_{K,m,m'} t_{k} c_{m'k}^{\dagger}d_{m} + h.c.$. 
In addition to the DQD case, the coupling matrix elements $\nu_{K,m,m'}$ obtain 
an additional index, encoding spin-flip processes upon tunneling of an electron with spin $m$ into lead $K$. 
The role of the hopping and the spin-flip terms is similar in both systems. They mediate a coherent 
superposition of the physical degrees of freedom when $\epsilon_{\text{a}/\uparrow}\approx\epsilon_{\text{b}/\downarrow}$, 
giving rise to sizeable coherences \cite{Hartle2013b,Hartle2014} and, thus, an exchange interaction 
that quenches the current suppression in a rather narrow range of voltages (see below). 
Please note that the left and right degrees of freedom of the DQD can be considered as a pseudo-spin degree of 
freedom that allows a mapping to a spinful Anderson impurity coupled to a single polarized 
electrode \cite{Kashcheyevs2007}. This mapping, however, does not include a second electrode with a different polarization and/or 
chemical potential such that, on a formal level, the DQD and SV case are similar but definitely not identical.

\emph{Transport theory:} We determine the transport properties of the two systems 
using the HQME technique \cite{Jin2008,Hartle2013b}, where we solve the coupled equations of motion  
\begin{eqnarray} 
\label{hierarcheom}
\partial_{t}\rho_{j_{1}..j_{\kappa}}^{(\kappa)}(t) & = & 
-i\left[H_{\text{Imp}}^{A},\rho_{j_{1}..j_{\kappa}}^{(\kappa)}(t)\right] 
- \hspace{-0.1cm}\sum_{\lambda\in\{1..\kappa\}}\hspace{-0.1cm}\omega_{K_{\lambda},p_{\lambda}}^{s_{\lambda}}\rho_{j_{1}..j_{\kappa}}^{(\kappa)}(t)\nonumber\\
 &&\hspace{-1.5cm}+\sum_{\lambda}(-1)^{\kappa-\lambda} 
 \eta_{K_{\lambda},p_{\lambda}}^{s_{\lambda}} d_{m_{\lambda}}^{s_{\lambda}} \rho_{j_{1}..j_{\kappa}/j_{\lambda}}^{(\kappa-1)}(t) \nonumber\\ 
 &&\hspace{-1.5cm}+\sum_{\lambda}(-1)^{\lambda}  
 \eta_{K_{\lambda},p_{\lambda}}^{\overline{s}_{\lambda},*}\rho_{j_{1}..j_{\kappa}/j_{\lambda}}^{(\kappa-1)}(t)d_{m_{\lambda}}^{s_{\lambda}}\nonumber \\
 &&\hspace{-1.5cm}-\hspace{-0.2cm}\sum_{j_{\kappa+1},m'}\hspace{-0.2cm} \Xi^{s_{\kappa+1}}_{K_{\kappa+1},m'm_{\kappa+1}}
 \left[ d_{m'}^{\overline{s}_{\kappa+1}}, \rho_{j_{1}..j_{\kappa+1}}^{(\kappa+1)}(t) \right]_{(-1)^{\kappa+1}},
\end{eqnarray}
starting from a product initial state. 
The density matrix of the impurity enters as $\rho^{(0)}(t)=\rho(t)$. 
The operators $\rho_{j_{1}..j_{\kappa}}^{(\kappa\geq1)}(t)$ describe its 
dynamics due to the coupling to the electrodes. 
They involve superindices $j=(K,m,s,p)$ with a lead index $K$, a level index $m$ and an index $s\in\{+,-\}$ 
corresponding to creation and annihilation processes. The index $p$ is related to a sum-over-poles decomposition 
of the hybridization and distribution function in the leads. 
The electrical current 
$I_{K}= \sum_{K,mm',p}\Xi_{K,mm'}^{+}\text{Tr}\left[\rho_{K,m,+,p}^{(1)}(t)d_{m'}\right] + h.c.$ ($K\in\text{L,R}$) 
is given by the first tier operators. 
Details of the derivation and explicit expressions for the DQD system 
can be found in Refs.\ \onlinecite{Hartle2013b,Hartle2014}. Here, we present a 
generalization of the HQME formalism to the SV case, where the coupling factors $\Xi_{K,mm'}^{s}$ 
are given by $\nu^{s}_{K,m}\nu^{\overline{s}}_{K,m'}/\gamma$ 
and $(\sum_{\tilde{m}}\nu^{s}_{K,\tilde{m},m}\nu^{\overline{s}}_{K,\tilde{m},m'})/\gamma$ 
in the DQD and SV scenario, respectively, and $\gamma$ is the band width. 
Note that the HQME (\ref{hierarcheom}) are time-local and, therefore, the only method that can be used so far to obtain  
exact results on the long-lived dynamics in these systems 
($t\sim10^{2}$--$10^{3}/\Gamma$, where $\Gamma$ denotes the hybridization strength, cf.\ Ref.\ \cite{Hartle2014}). 

The hierarchy (\ref{hierarcheom}) is infinite, but can be truncated by discarding all terms 
that belong to a specific tier $\kappa$ and higher. This represents an approximation of the time evolution operator 
up to $(\kappa-1)$th order in the hybridization strength $\Gamma$. 
Another, systematic truncation scheme is based on assigning each operator $\rho_{j_{1}..j_{\kappa}}^{(\kappa)}(t)$ 
the importance value \cite{Hartle2013b} 
\begin{eqnarray}
\left\vert \prod_{\lambda=1..\kappa}\frac{\eta_{K_{\lambda},m_{\lambda},p_{\lambda}}^{s_{\lambda}} 
\sum_{m'} \Xi_{K,m_\lambda m'}^{s_\lambda} }{ 2\sum_{\lambda'=1..\lambda}\text{Re}[\omega_{K_{\lambda'},p_{\lambda'}}^{s_{\lambda'}}]
\text{Re}[\omega_{K_{\lambda},p_{\lambda}}^{s_{\lambda}}] 
} \right\vert ,\label{ampli}
\end{eqnarray}
and including only those terms with a value exceeding a certain threshold. 
This reduces the numerical complexity to a practical level. Converged or exact results 
can be obtained for high enough temperatures (typically $T\gtrsim T_K$, see Ref.\ \cite{Hartle2015}) 
while, at lower temperatures, convergence becomes prohibitive. 
Note that (\ref{ampli}) is an extension to the SV case.

\begin{figure}[b]
\begin{tabular}{ll}
\resizebox{\newwidthII}{\newheightII}{ 
\includegraphics{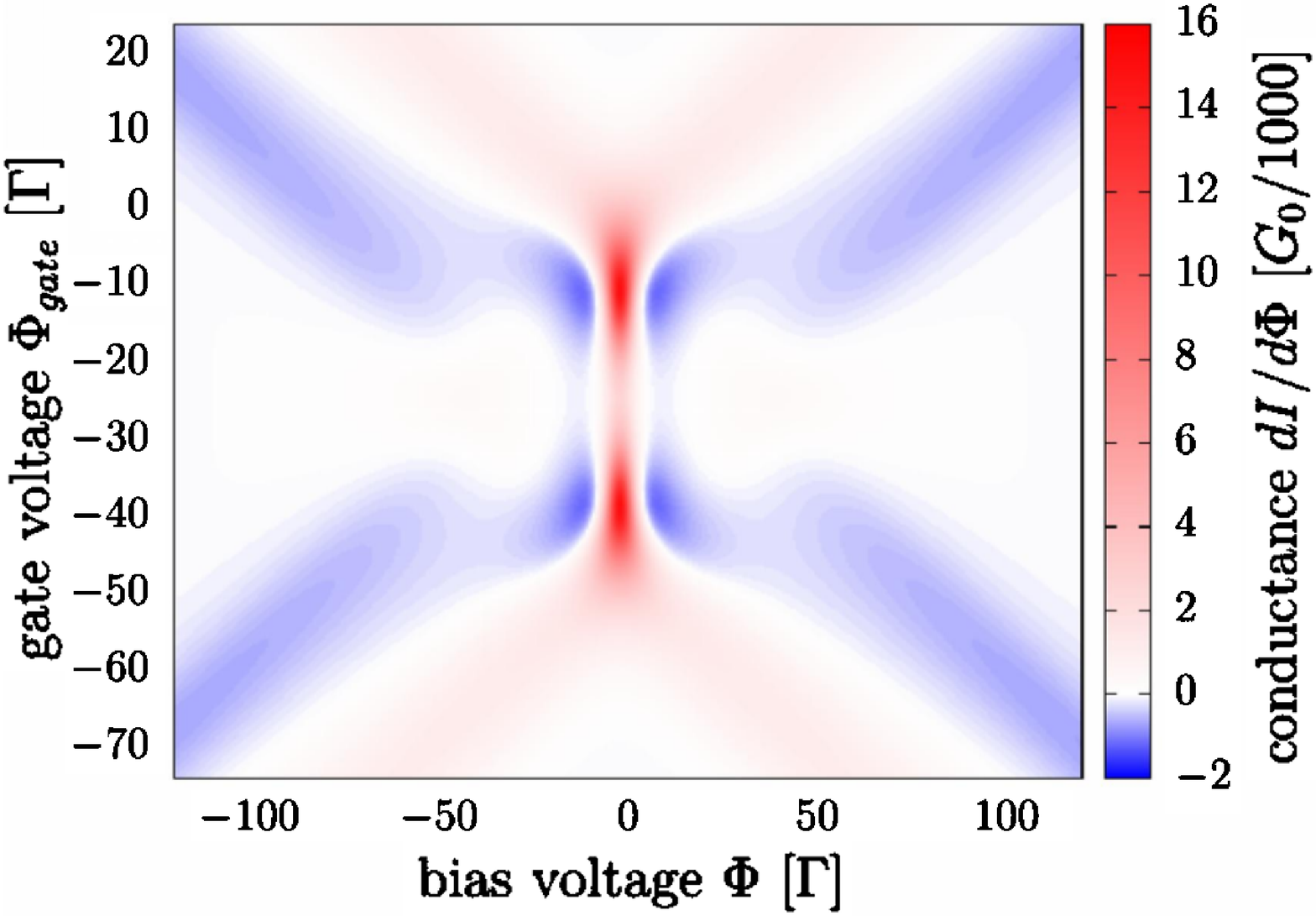} 
}& 
\resizebox{\newwidthII}{\newheightII}{ 
\includegraphics{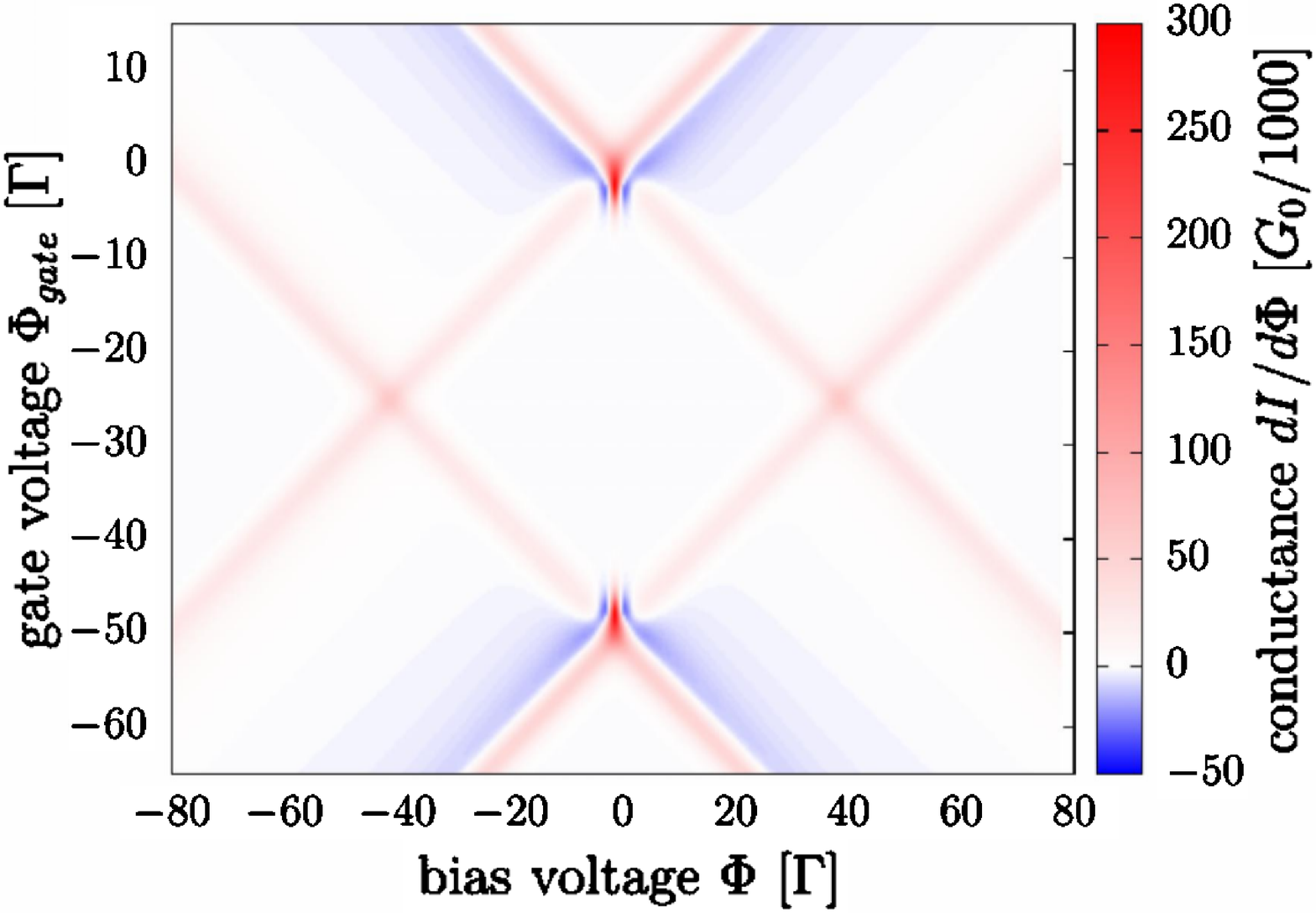} 
}\\
\resizebox{\newwidthII}{\newheightII}{ 
\includegraphics{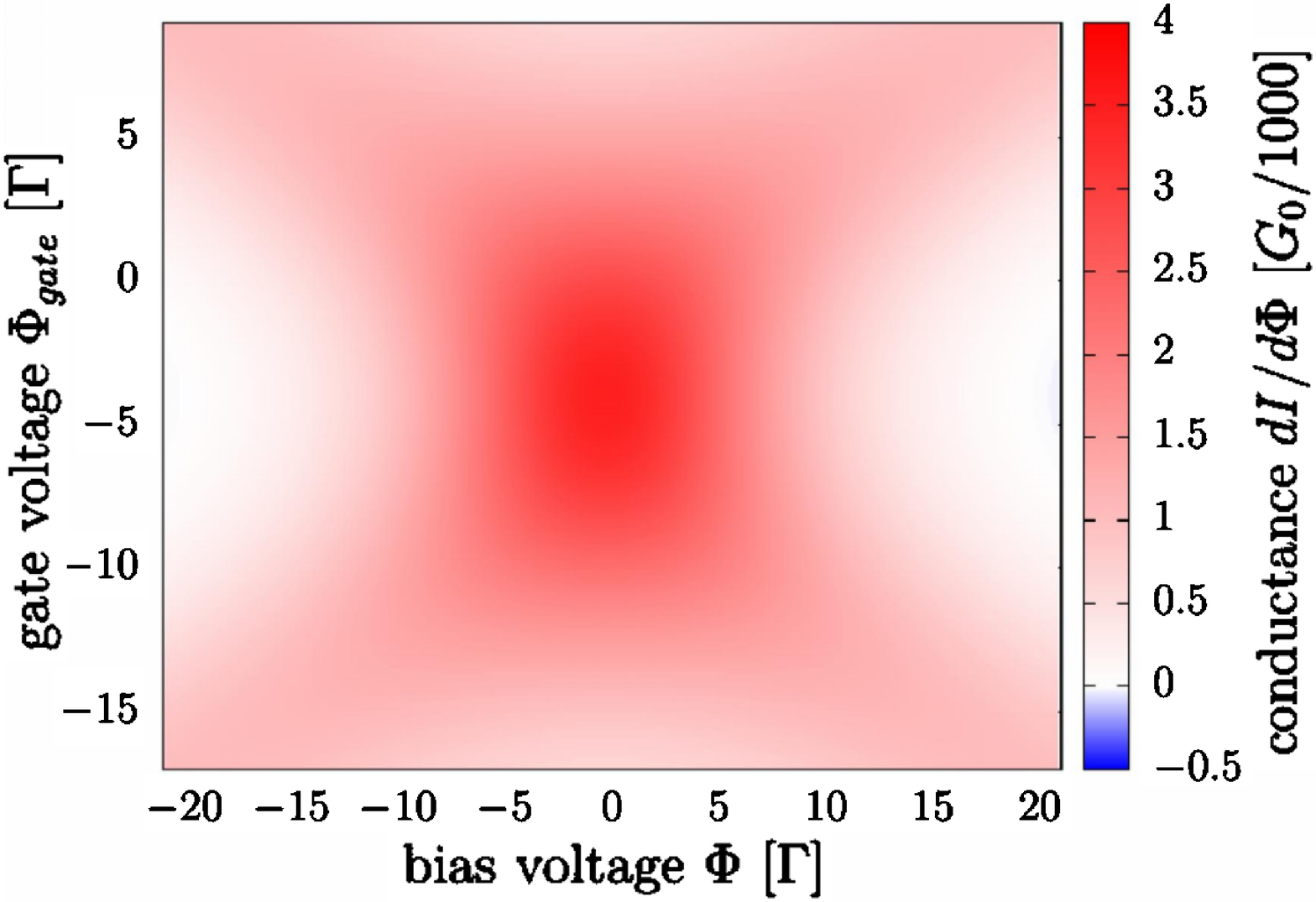} 
}& 
\resizebox{\newwidthII}{\newheightII}{ 
\includegraphics{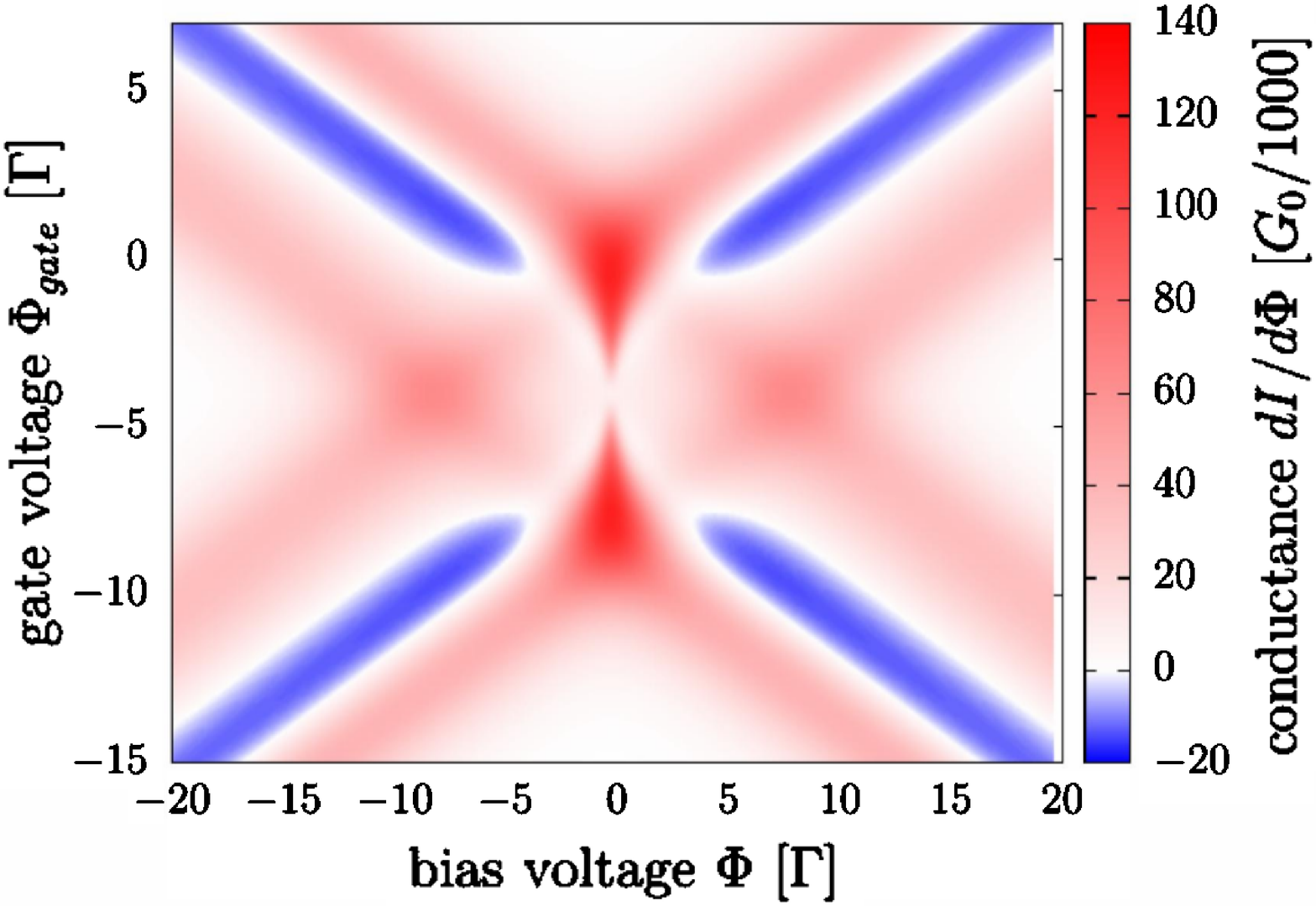} 
}\\
\end{tabular}\caption{\label{firstordermapssym} Conductance maps for symmetric coupling to electrodes, 
$\Gamma_{\text{L}}=\Gamma_{\text{R}}=\Gamma$. The left and right panels refer to the DQD and SV case, respectively. 
The upper panels are computed with $U=50\Gamma$; the lower panels with $U=8\Gamma$. For even lower U, 
the maps are similar to the one on the lower left side.}
\end{figure}

\newpage
We also present results obtained from the well established Born-Markov master equation (BMME) 
\cite{May02,Mitra04,Lehmann04,Harbola2006,Semmelhack,Siddiqui,Timm08,Hartle09,Hartle2010b}
\begin{eqnarray}
\label{redmatEOM}
\partial_t \rho(t) &=&  - i \left[ H^{A}_{\text{Imp}} , \rho(t) \right] \\
&&\hspace{-1cm}- \int_{0}^{t}\text{d}\tau\,
\text{tr}_{\text{L+R}}\bigl\lbrace\bigl[\,H^{A}_{\text{Imp,L+R}},\left[\,H^{A}_{\text{Imp,L+R}}(\tau),\rho(t)  
\rho_{\text{L+R}}^{\text{equ.}}\right]\bigr]\rbrace, \nonumber
\end{eqnarray}
where $\rho_{\text{L+R}}^{\text{equ.}}$ denotes the equilibrium density matrix of the leads. 
The level of approximation is the same as a truncation of (\ref{hierarcheom}) at the first tier, 
including the Markov approximation 
$\rho(t-\tau)\approx\text{e}^{i H_{\text{Imp}}^{A}\tau} \rho(t) \text{e}^{-i H_{\text{Imp}}^{A}\tau}$.

\emph{Results:} 
We consider different parameter regimes of the two systems in order to cover 
different facets of the phenomenon, in particular slightly higher temperatures in the DQD, $T=4.5\Gamma$, 
as compared to the SV case, $T=\Gamma$ 
(where we use the hybridization strength $\Gamma=2\pi\text{max}[\Xi_{K,mm}]$ as 
the unit of energy). 
The electrodes are modeled in both systems by Lorentzian bands, that is  
$\sum_{k\in K} \vert t_k\vert^2 \delta(\epsilon-\epsilon_{k/mk}) = \gamma / ( (\omega-\mu_K)^2 + \gamma^2 )$. 
Note that this choice is not crucial, as we use $\gamma\gtrsim50\Gamma$ to avoid band edge effects. 
The bias voltage $\Phi$ enters via $\mu_{\text{L/R}}=\pm \Phi/2$, 
the gate voltage via a simple shift of the degenerate single-particle levels, $\epsilon_{m}=\Phi_{\text{gate}}$. 
The current is strongly suppressed in the DQD case by a weak coupling 
between the dots, $t_{\text{hopp}}\approx \Gamma/11$, and in the SV system  
by almost fully (99\%) polarized electrodes 
and an almost antiparallel configuration of the 
respective polarization vectors ($\alpha=\pm0.05\pi$, cf.\ right panel of \text{Fig.\ \ref{qdfig}}).

We introduce the phenomenon by discussing the conductance maps depicted in Fig.\ \ref{firstordermapssym}. 
The four panels are organized such that the left/right column 
represents the DQD/SV case, while the top row corresponds to calculations with strong electron-electron interactions, 
$U=50\Gamma$, and the bottom row to an intermediate coupling strength, $U=8\Gamma$. All plots show a prominent peak 
around zero bias $\Phi\approx0$ that extends from $\Phi_{\text{gate}}=0$ to $\Phi_{\text{gate}}=-U$. It has a maximum 
at a gate voltage $-U/2<\Phi_{\text{gate}}<0$ (or $-U<\Phi_{\text{gate}}<-U/2$). 
The position of the peak is very similar to a 'Kondo peak', except that the associated conductance value $G$ 
is much lower than the conductance quantum, $G\ll G_0$. A technical reason why this can't be a Kondo resonance is that 
these calculations are obtained from Born-Markov theory. Similar conductance maps have been obtained using first 
(cf.\ Fig.\ 11 of Ref.\ \onlinecite{Hartle2013b}) 
and second order perturbation theory (cf.\ Fig.\ 7 of Ref.\ \onlinecite{Hell2014}). 
Can we trust the approximate results, in particular as 
first order processes are strongly suppressed at these voltages and 
higher order effects dominant?

\begin{figure}
\begin{tabular}{ll}
\resizebox{\newwidth}{\newheight}{\includegraphics{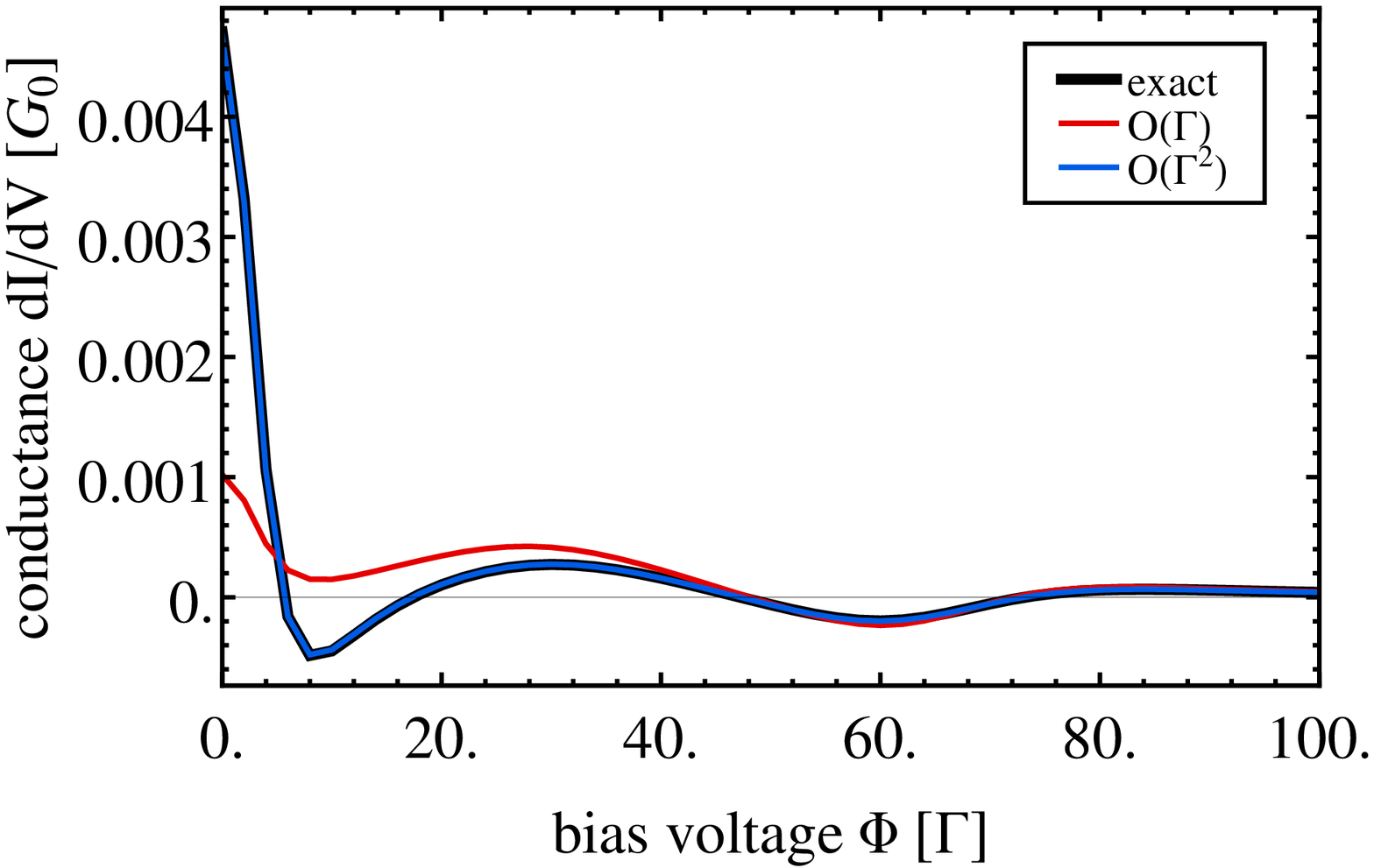}} & 
\resizebox{\newwidth}{\newheight}{\includegraphics{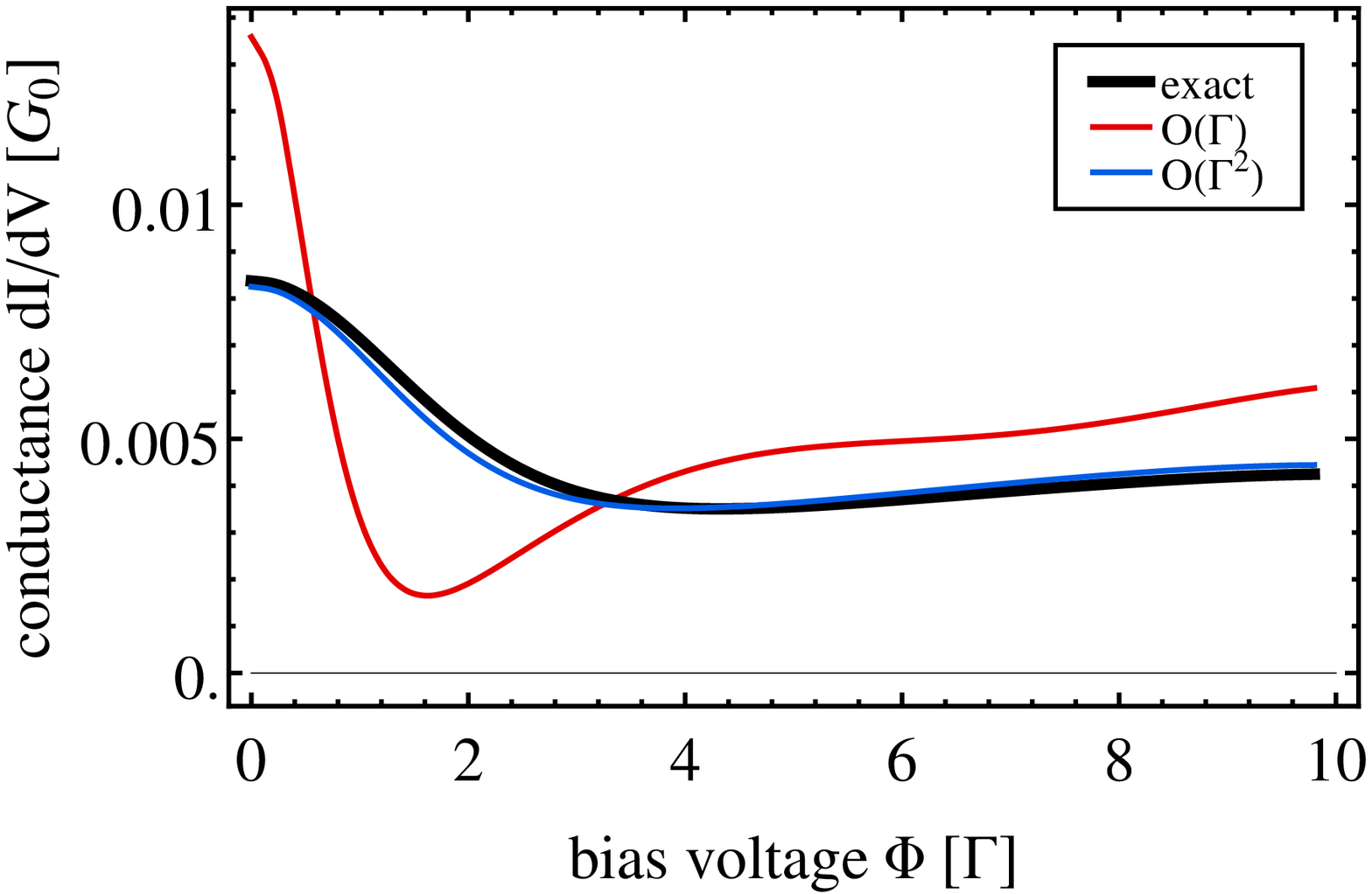}} \\
\end{tabular}
\caption{\label{effectofhigherordershifted} Conductance-voltage characteristics of the DQD with $U=50\Gamma$ 
and $\epsilon_{m}=-0.3U$ (left panel) and the SV with $U=8\Gamma$ and $\epsilon_{m}=-U/4$ (right panel). }
\end{figure}

We verify the validity of previous results by the conductance-voltage characteristics that are shown 
in Fig.\ \ref{effectofhigherordershifted}. The three lines correspond to the first (red line), second order (blue line) 
and a fully converged result (black line, which includes terms up to $\mathcal{O}(\Gamma^4)$), 
where the left panel represents the DQD case with strong 
and the right panel the SV case with weaker electron-electron interactions. The gate voltage $\Phi_{\text{gate}}$ 
is fixed to a value between $0$ and $-U/2$ where the conductance peak appears most pronounced 
(cf.\ dashed black line in the left panel of Fig.\ \ref{qdfig}). 
Overall, the phenomenon turns out to be robust. 
The second order result is very close to the exact one. 
Differences between the second order and exact results are significant at lower temperatures 
(compare SV and DQD case with $T=\Gamma$ and $T=4.5\Gamma$, respectively). 

\begin{SCfigure}
\resizebox{\newwidth}{\newheight}{\includegraphics{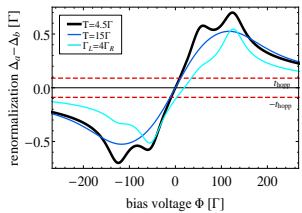}} 
\caption{\label{shiftplot} 
Difference of the interaction-induced renormalizations $\Delta_{\text{a}}-\Delta_{\text{b}}$ (cf.\ Eq.\ (\ref{intindrenorm})). 
The red dashed lines mark the narrow range of voltages where $\vert\Delta_{\text{a}}-\Delta_{\text{b}}\vert<t_{\text{hopp}}$. }
\end{SCfigure}

The origin of the zero bias peak is a strong suppression of the current flow in both systems 
and interaction-induced renormalization effects \cite{Martinek2003}. 
They are included in the principal value terms of the time integrals in (\ref{redmatEOM}). 
In the DQD case, they add to the energy levels $\epsilon_{a/b}\rightarrow\epsilon_{a/b}+\Delta_{a/b}$ 
and can be estimated by \cite{Wunsch2005,Hartle2014} 
\begin{eqnarray}
\label{intindrenorm}
 \Delta_{a/b} = \phi_{\text{a/b}}(\epsilon_{a/b},\mu_{\text{L/R}}) - \phi_{\text{a/b}}(\epsilon_{a/b}+U,\mu_{\text{L/R}}), 
\end{eqnarray}
with $\phi_{\text{a/b}}(x,\mu) = 
\frac{\Gamma_{\text{L/R}}}{2\pi}\text{Re}\left[ \Psi\left( \frac{1}{2} + \frac{i (x-\mu)}{2\pi T} \right) \right]$ 
and the digamma function $\Psi(x)$. Clearly, if they are neglected, the sharp peak features do not appear (cf.\ Figs.\ 11 and 12 of Ref.\ \cite{Hartle2013b}). 
For bias voltages $\Phi\neq0$, the two levels are thus shifted 
in different directions (see Fig.\ \ref{shiftplot}). This is important in situations where 
$\epsilon_m\ll\mu_{\text{L/R}}\ll\epsilon_m+U$, \emph{i.e.}\ where the dots are populated by one electron. 
The electron is additionally trapped when $\epsilon_\text{a}=\epsilon_{\text{b}}$ such that it can coherently 
oscillate between the two dots. This trapping is quenched when the levels are disaligned due to the 
interaction-induced renormalizations $\Delta_{\text{a/b}}$. 
The corresponding range of voltages is rather narrow and, using the uncertainty principle 
$\Delta E \Delta t = 1/2$ with $\Delta t=\pi/t_{\text{hopp}}$, 
can be estimated by $\vert\Delta_{a}-\Delta_{b}\vert\sim t_{\text{hopp}}$ (see Fig.\ \ref{shiftplot}). 
The renormalizations $\Delta_{\text{a/b}}$ originate from coherent exchange processes with the electrodes and become 
effective only in the presence of electron-electron interactions. They translate, inter alia, to a coupling between the 
eigenstates of the DQD $\sim \Delta_{\text{a}}-\Delta_{\text{b}}$ \cite{Hartle2013b}. The corresponding degrees of freedom in the SV case 
are the spin degrees of freedom, where the interaction-induced coupling 
quenches the current suppression due to the almost antiparallel polarization of the electrodes. 
This coupling is the exchange interaction mentioned above. 
Its basics can be understood in terms of first-order arguments 
but second and beyond-second order effects are apparently important.  
The coupling interpretation shows that it always leads to an enhanced conductance 
at low voltages. Otherwise, it can lead to a decrease and even 
negative differential conductance \cite{Wunsch2005,Hettler2007,Trocha2009,Donarini2010} 
or gate-dependent line shapes \cite{Paaske2008,Begemann2010}.

\begin{figure}
\begin{tabular}{ll}
\resizebox{\newwidth}{\newheight}{\includegraphics{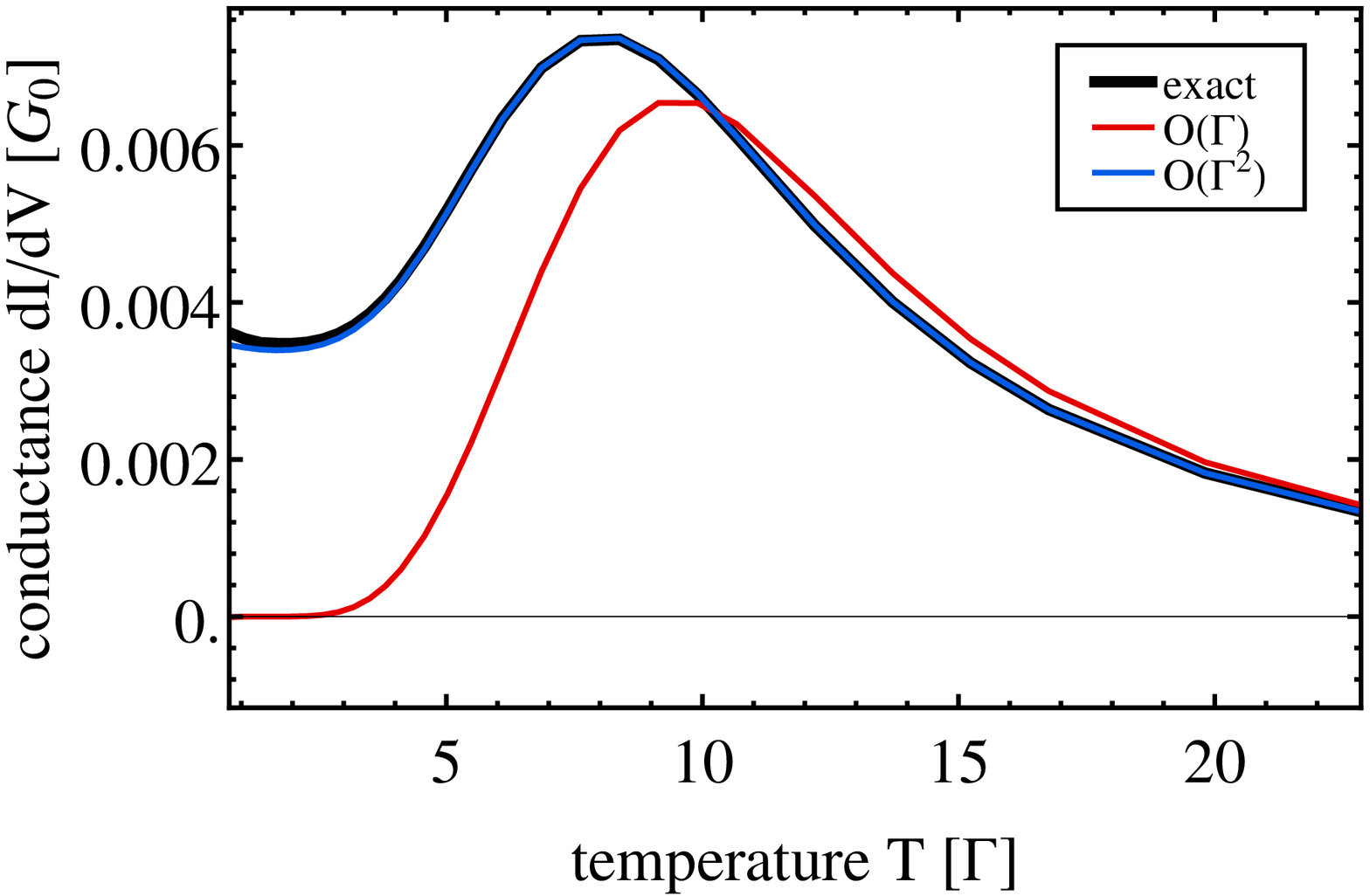} }& 
\resizebox{\newwidth}{\newheight}{\includegraphics{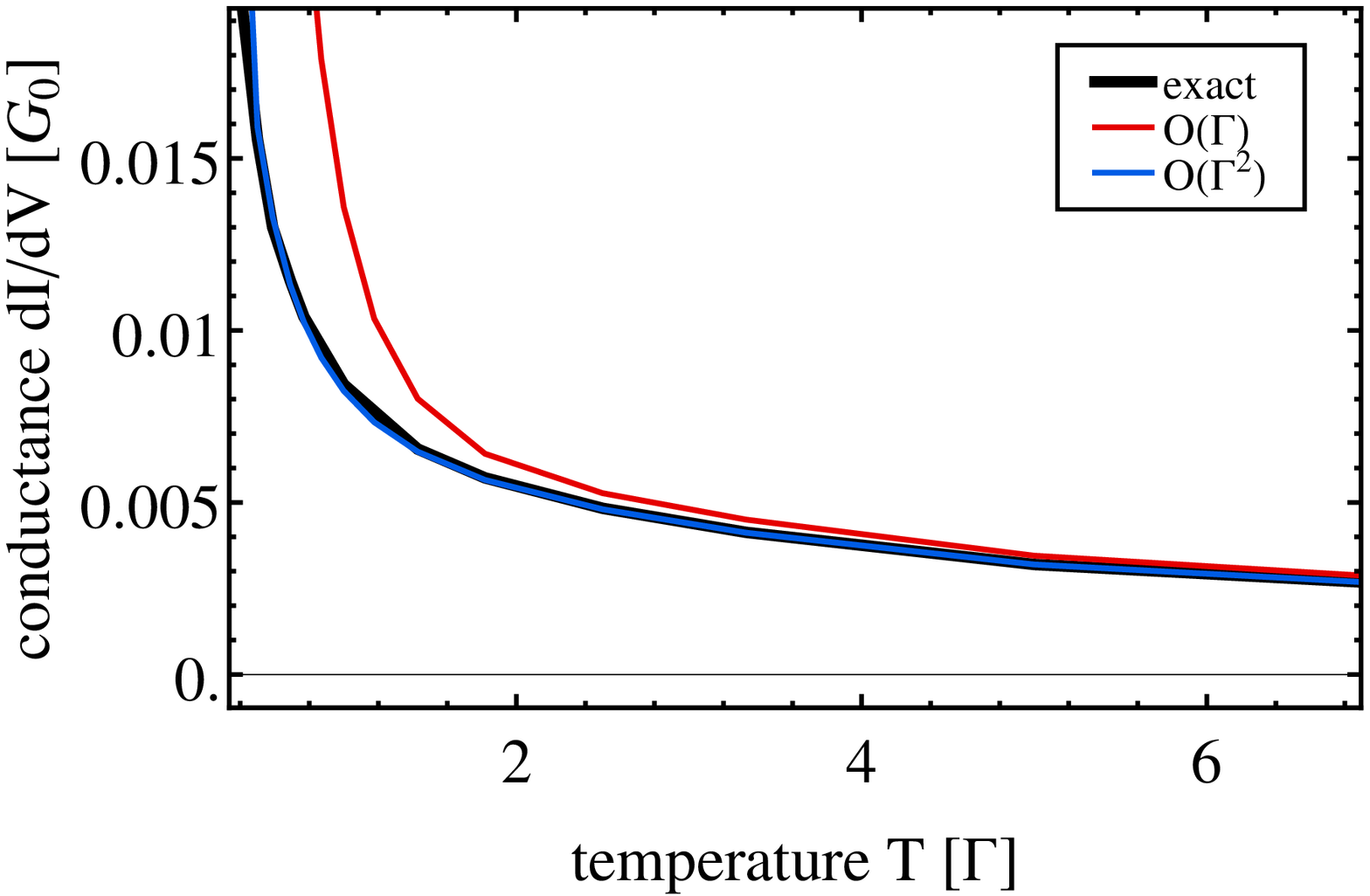} }\\
\end{tabular}
\caption{\label{tdependenceshifted} Zero-bias conductance of the DQD with $U=50\Gamma$ 
and $\epsilon_{m}=-0.3U$ (left panel) and the SV with $U=8\Gamma$ and $\epsilon_{m}=-U/4$ (right panel) 
as a function of temperature $T$. }
\end{figure}

\begin{figure}
\begin{tabular}{ll}
\resizebox{\newwidth}{\newheight}{\includegraphics{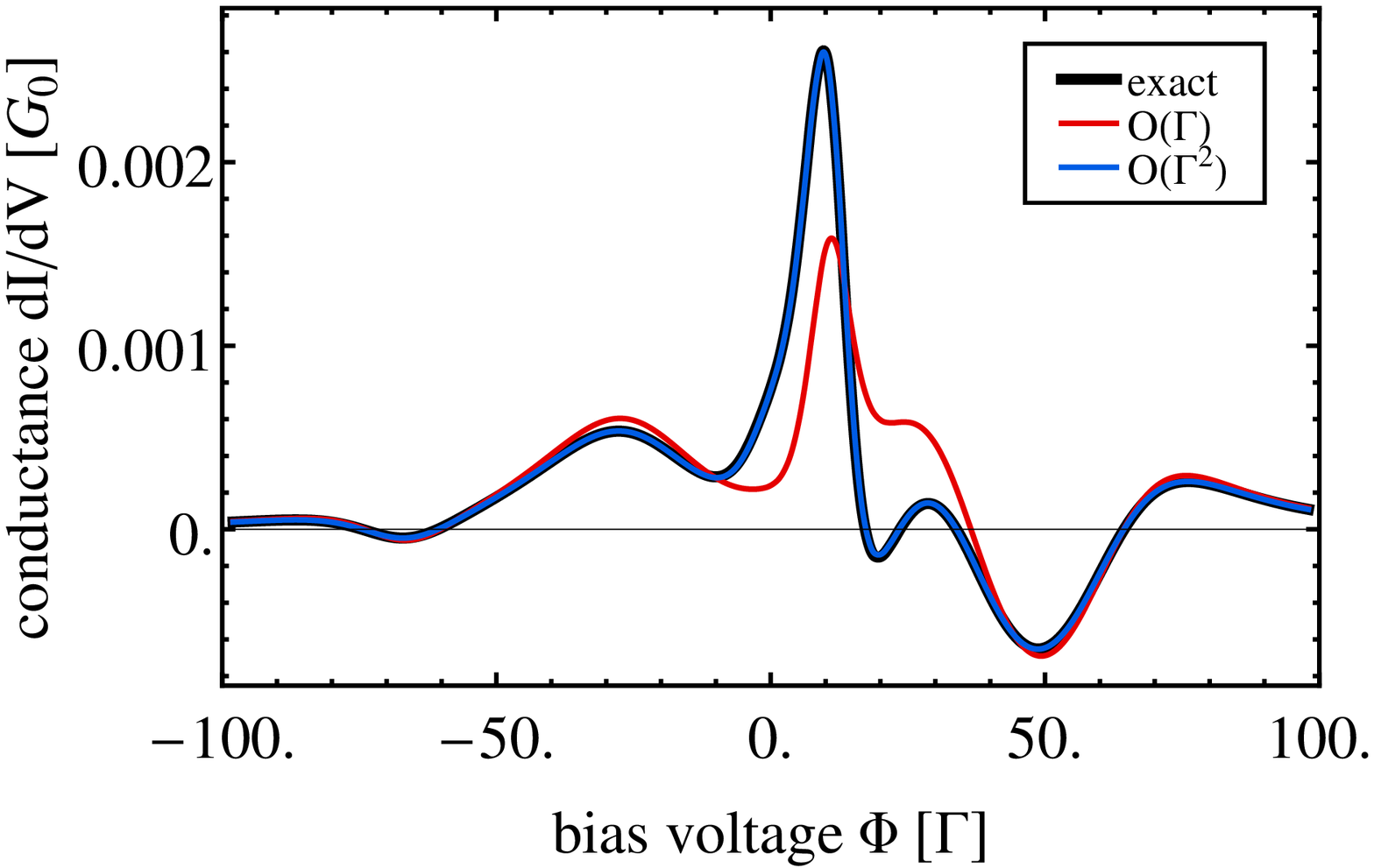}} & 
\resizebox{\newwidth}{\newheight}{\includegraphics{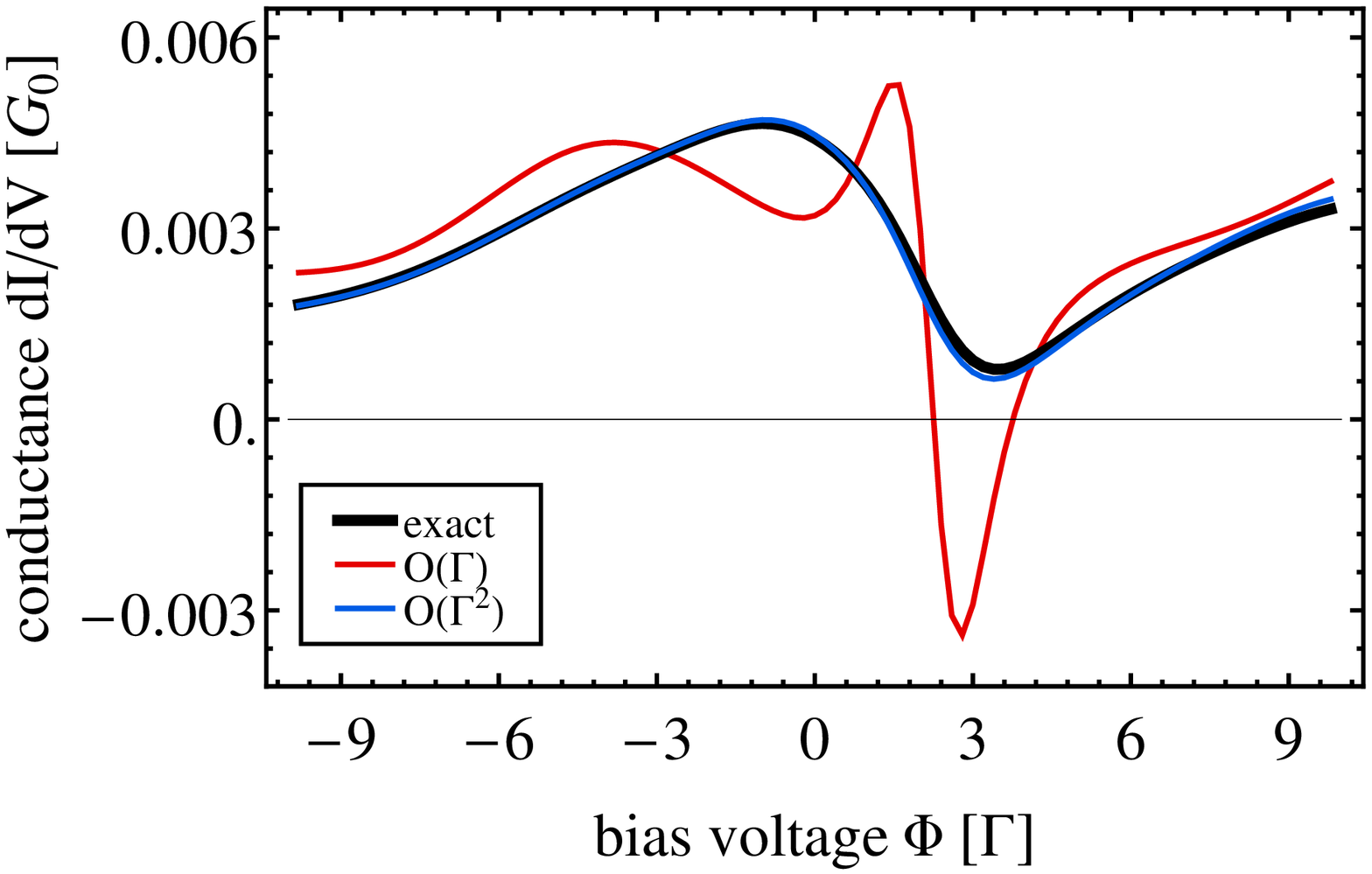} } \\
\end{tabular}
\caption{\label{asymshifted} 
Conductance-voltage characteristics of the DQD with $U=50\Gamma$ 
and $\epsilon_{m}=-0.3U$ (left panel) and the SV with $U=8\Gamma$ and $\epsilon_{m}=-U/4$ (right panel) 
for an asymmetric coupling to the electrodes, $\Gamma_{\text{L}}=\Gamma=4\Gamma_{\text{R}}$.}
\end{figure}

Now, we would like to explain how the sharp peaks can be distinguished from Kondo peaks, since 
the unitary limit $G=G_0$ is not accessible in all experimental setups. Thus, 
evidence is typically derived from the temperature dependence of the conductance, which we show  
in Fig.\ \ref{tdependenceshifted}. 
The DQD case (left panel) shows a non-monotonic peak structure at intermediate temperatures 
that is followed by a power-law decay $\sim1/T^{2}$ (which converts to a well-known $1/T$ scaling 
when thermal broadening becomes dominant; see lower left panel of Fig.\ \ref{firstordermapssym}). 
The same scaling is also observed in the SV case (right panel). 
This behavior results from the non-monotonic temperature dependence of the 
interaction-induced renormalizations, that is $\Delta_a - \Delta_b \stackrel{T\rightarrow0}{=}u+vT^2$ 
and $\Delta_a - \Delta_b \stackrel{T\rightarrow\infty}{\sim}\Phi/T^2$, 
including a peak at intermediate temperatures which is also seen in the conductance-temperature characteristics. 
For weaker $U$, it is shifted to 
lower temperatures and, therefore, not visible in the weak coupling SV case (right panel of Fig.\ \ref{tdependenceshifted}). 
This analytic finding is in clear contrast to a Kondo peak, where we expect a monotonic decay and a different scaling 
behavior \cite{Hewson93,Goldhaber1998b}.  
Note that the conductance maxima (red spots in the upper panels of 
\text{Fig.\ \ref{firstordermapssym}}) also show an interesting temperature dependence. 
In the DQD case with $U=50\Gamma$, they are shifted towards the charge-symmetric point when 
increasing temperature (data not shown). This is in clear contrast to Kondo peaks, 
but we also observe the opposite behavior such that a clear distinction is not possible.

The cleanest distinction can be made in systems where the coupling to the electrodes is asymmetric, 
$\Gamma_{\text{L}}\neq\Gamma_{\text{R}}$. 
The conductance-voltage characteristics of such scenarios are shown in Fig.\ \ref{asymshifted}. 
In contrast to symmetric coupling, the sharp conductance peaks no longer appear at zero 
but at non-zero bias voltages \cite{Hell2014}. This can be readily seen by the dependence of 
the interaction-induced renormalizations (\ref{intindrenorm}) on the hybridization strengths, $\Delta_{\text{a/b}}\sim\Gamma_{\text{L/R}}$ 
(see turquoise line in Fig.\ \ref{shiftplot}). 
For $\Gamma_{\text{L}}>\Gamma_{\text{R}}$, the root of the difference $\Delta_{\text{a}}-\Delta_{\text{b}}$ is shifted from 
$\Phi=0$ to a non-zero, positive bias voltage, which corresponds to the peak position in the 
conductance-voltage characteristics. The maximal shift is $\Phi=\pm(2\epsilon_0+U)$, which can be derived from 
the condition $\Delta_{\text{a/b}}(\Phi)=0$ in the limit $\Gamma_{\text{L/R}}\rightarrow0$. 
The peak structure gets suppressed as $\Gamma_{\text{L/R}}\rightarrow0$ 
and vanishes faster including higher order effects 
(cf.\ right panel of Fig.\ \ref{asymshifted}). Note that a similar behavior emerges 
for detuned levels, where $\epsilon_{\text{a}}-\epsilon_{\text{b}}=0.4\Gamma$ corresponds to 
$\Gamma_{\text{L}}=\Gamma=4\Gamma_{\text{R}}$.

\emph{Conclusions:} We studied sharp peaks in the conductance-voltage characteristics 
of double quantum dot and spin-valve systems with (quasi-)degenerate single-particle levels. 
We showed that the peaks can be understood in both systems on the same 
footing, namely in terms of exchange interactions that quench a strong, inherent current suppression. 
In symmetric systems, the peaks are located at the same voltages as Kondo peaks but can be clearly 
distinguished by their $1/T^2$, sometimes even non-monotonic temperature dependence. 
In asymmetric or detuned systems, the distinction is even easier, because the peaks occur at non-zero bias voltages 
and one bias polarity \cite{Hell2014}. 
Our analysis is based on the hierarchical quantum master equation technique, 
which allows us to compare first and second order results with numerically converged, exact results. 
Thus, we were able to demonstrate the robustness of the phenomenon and the role of higher order effects.  
We also used results derived from Born-Markov theory, demonstrating the importance of 
principal value terms in the presence of quasidegenerate energy levels. 
Our study underlines the importance of an exact description of exchange interactions. This can be even more 
relevant in transport through more complex systems such as, for example, multi-dot systems or 
molecular junctions where quasidegeneracies are even more important. In contrast to Kondo peaks, 
the peaks described here are more robust with respect to bias voltage and temperature. 
This is beneficial from both an experimental and a technological point of view.

\emph{Acknowledgement:} 
We thank A.J.\ Millis, J.\ Okamoto, M.\ Wegewijs, J.\ K\"onig, 
P.\ Wang, K.\ Sch\"onhammer 
and S.\ Kehrein for helpful comments and discussions. 
RH is supported by Deutsche Forschungsgemeinschaft (DFG) under grant No.\ HA 7380/1-1.


\begin{thebibliography}{45}%
\makeatletter
\providecommand \@ifxundefined [1]{%
 \@ifx{#1\undefined}
}%
\providecommand \@ifnum [1]{%
 \ifnum #1\expandafter \@firstoftwo
 \else \expandafter \@secondoftwo
 \fi
}%
\providecommand \@ifx [1]{%
 \ifx #1\expandafter \@firstoftwo
 \else \expandafter \@secondoftwo
 \fi
}%
\providecommand \natexlab [1]{#1}%
\providecommand \enquote  [1]{``#1''}%
\providecommand \bibnamefont  [1]{#1}%
\providecommand \bibfnamefont [1]{#1}%
\providecommand \citenamefont [1]{#1}%
\providecommand \href@noop [0]{\@secondoftwo}%
\providecommand \href [0]{\begingroup \@sanitize@url \@href}%
\providecommand \@href[1]{\@@startlink{#1}\@@href}%
\providecommand \@@href[1]{\endgroup#1\@@endlink}%
\providecommand \@sanitize@url [0]{\catcode `\\12\catcode `\$12\catcode
  `\&12\catcode `\#12\catcode `\^12\catcode `\_12\catcode `\%12\relax}%
\providecommand \@@startlink[1]{}%
\providecommand \@@endlink[0]{}%
\providecommand \url  [0]{\begingroup\@sanitize@url \@url }%
\providecommand \@url [1]{\endgroup\@href {#1}{\urlprefix }}%
\providecommand \urlprefix  [0]{URL }%
\providecommand \Eprint [0]{\href }%
\providecommand \doibase [0]{http://dx.doi.org/}%
\providecommand \selectlanguage [0]{\@gobble}%
\providecommand \bibinfo  [0]{\@secondoftwo}%
\providecommand \bibfield  [0]{\@secondoftwo}%
\providecommand \translation [1]{[#1]}%
\providecommand \BibitemOpen [0]{}%
\providecommand \bibitemStop [0]{}%
\providecommand \bibitemNoStop [0]{.\EOS\space}%
\providecommand \EOS [0]{\spacefactor3000\relax}%
\providecommand \BibitemShut  [1]{\csname bibitem#1\endcsname}%
\let\auto@bib@innerbib\@empty
\bibitem [{\citenamefont {Heisenberg}(1926)}]{Heisenberg1926}%
  \BibitemOpen
  \bibfield  {author} {\bibinfo {author} {\bibfnamefont {W.}~\bibnamefont
  {Heisenberg}},\ }\href@noop {} {\bibfield  {journal} {\bibinfo  {journal} {Z.
  Physik}\ }\textbf {\bibinfo {volume} {38}},\ \bibinfo {pages} {411} (\bibinfo
  {year} {1926})}\BibitemShut {NoStop}%
\bibitem [{\citenamefont {Dirac}(1926)}]{Dirac1926}%
  \BibitemOpen
  \bibfield  {author} {\bibinfo {author} {\bibfnamefont {P.~A.~M.}\
  \bibnamefont {Dirac}},\ }\href@noop {} {\bibfield  {journal} {\bibinfo
  {journal} {Proceedings of the Royal Society of London. Series A, Containing
  Papers of a Mathematical and Physical Character}\ }\textbf {\bibinfo {volume}
  {112}},\ \bibinfo {pages} {661} (\bibinfo {year} {1926})}\BibitemShut
  {NoStop}%
\bibitem [{\citenamefont {Goodenough}(1966)}]{Goodenough1966}%
  \BibitemOpen
  \bibfield  {author} {\bibinfo {author} {\bibfnamefont {J.~B.}\ \bibnamefont
  {Goodenough}},\ }\href@noop {} {\emph {\bibinfo {title} {Magnetism and the
  Chemical Bond}}}\ (\bibinfo  {publisher} {Interscience Publishers},\ \bibinfo
  {address} {New York},\ \bibinfo {year} {1966})\BibitemShut {NoStop}%
\bibitem [{\citenamefont {White}(2007)}]{White2007}%
  \BibitemOpen
  \bibfield  {author} {\bibinfo {author} {\bibfnamefont {R.~W.}\ \bibnamefont
  {White}},\ }\href@noop {} {\emph {\bibinfo {title} {Quantum Theory of
  Magnetism: Magnetic Properties of Materials}}}\ (\bibinfo  {publisher}
  {Springer},\ \bibinfo {address} {Berlin},\ \bibinfo {year}
  {2007})\BibitemShut {NoStop}%
\bibitem [{\citenamefont {Kohn}\ and\ \citenamefont
  {Sham}(1965)}]{KohnSham1965}%
  \BibitemOpen
  \bibfield  {author} {\bibinfo {author} {\bibfnamefont {W.}~\bibnamefont
  {Kohn}}\ and\ \bibinfo {author} {\bibfnamefont {L.~J.}\ \bibnamefont
  {Sham}},\ }\href@noop {} {\bibfield  {journal} {\bibinfo  {journal} {Phys.
  Rev.}\ }\textbf {\bibinfo {volume} {140}},\ \bibinfo {pages} {A1133}
  (\bibinfo {year} {1965})}\BibitemShut {NoStop}%
\bibitem [{\citenamefont {Engel}\ and\ \citenamefont
  {Dreizler}(2011)}]{EngelDreizler2011}%
  \BibitemOpen
  \bibfield  {author} {\bibinfo {author} {\bibfnamefont {E.}~\bibnamefont
  {Engel}}\ and\ \bibinfo {author} {\bibfnamefont {R.~M.}\ \bibnamefont
  {Dreizler}},\ }\href@noop {} {\emph {\bibinfo {title} {Density Functional
  Theory}}}\ (\bibinfo  {publisher} {Springer},\ \bibinfo {address} {Berlin},\
  \bibinfo {year} {2011})\BibitemShut {NoStop}%
\bibitem [{\citenamefont {Kondo}(1964)}]{Kondo1964}%
  \BibitemOpen
  \bibfield  {author} {\bibinfo {author} {\bibfnamefont {J.}~\bibnamefont
  {Kondo}},\ }\href@noop {} {\bibfield  {journal} {\bibinfo  {journal} {Prog.
  Theor. Phys.}\ }\textbf {\bibinfo {volume} {32}},\ \bibinfo {pages} {37}
  (\bibinfo {year} {1964})}\BibitemShut {NoStop}%
\bibitem [{\citenamefont {Hewson}(1993)}]{Hewson93}%
  \BibitemOpen
  \bibfield  {author} {\bibinfo {author} {\bibfnamefont {A.~C.}\ \bibnamefont
  {Hewson}},\ }\href@noop {} {\emph {\bibinfo {title} {The Kondo problem to
  Heavy Fermions}}}\ (\bibinfo  {publisher} {Cambridge University Press},\
  \bibinfo {address} {Cambridge},\ \bibinfo {year} {1993})\BibitemShut
  {NoStop}%
\bibitem [{\citenamefont {{de Haas}}\ \emph {et~al.}(1934)\citenamefont {{de
  Haas}}, \citenamefont {{de Boer}},\ and\ \citenamefont {{van den
  Berg}}}]{deHaas1934}%
  \BibitemOpen
  \bibfield  {author} {\bibinfo {author} {\bibfnamefont {W.~J.}\ \bibnamefont
  {{de Haas}}}, \bibinfo {author} {\bibfnamefont {J.}~\bibnamefont {{de
  Boer}}}, \ and\ \bibinfo {author} {\bibfnamefont {G.~J.}\ \bibnamefont {{van
  den Berg}}},\ }\href@noop {} {\bibfield  {journal} {\bibinfo  {journal}
  {Physica (Utrecht)}\ }\textbf {\bibinfo {volume} {1}},\ \bibinfo {pages}
  {1115} (\bibinfo {year} {1934})}\BibitemShut {NoStop}%
\bibitem [{\citenamefont {Liang}\ \emph {et~al.}(2002)\citenamefont {Liang},
  \citenamefont {Shores}, \citenamefont {Bockrath}, \citenamefont {Long},\ and\
  \citenamefont {Park}}]{Liang02}%
  \BibitemOpen
  \bibfield  {author} {\bibinfo {author} {\bibfnamefont {W.}~\bibnamefont
  {Liang}}, \bibinfo {author} {\bibfnamefont {M.}~\bibnamefont {Shores}},
  \bibinfo {author} {\bibfnamefont {M.}~\bibnamefont {Bockrath}}, \bibinfo
  {author} {\bibfnamefont {J.}~\bibnamefont {Long}}, \ and\ \bibinfo {author}
  {\bibfnamefont {H.}~\bibnamefont {Park}},\ }\href@noop {} {\bibfield
  {journal} {\bibinfo  {journal} {Nature (London)}\ }\textbf {\bibinfo {volume}
  {417}},\ \bibinfo {pages} {725} (\bibinfo {year} {2002})}\BibitemShut
  {NoStop}%
\bibitem [{\citenamefont {Pasupathy}\ \emph {et~al.}(2004)\citenamefont
  {Pasupathy}, \citenamefont {Bialczak}, \citenamefont {Martinek},
  \citenamefont {Grose}, \citenamefont {Donev}, \citenamefont {McEuen},\ and\
  \citenamefont {Ralph}}]{Pasupathy2004}%
  \BibitemOpen
  \bibfield  {author} {\bibinfo {author} {\bibfnamefont {A.~N.}\ \bibnamefont
  {Pasupathy}}, \bibinfo {author} {\bibfnamefont {R.~C.}\ \bibnamefont
  {Bialczak}}, \bibinfo {author} {\bibfnamefont {J.}~\bibnamefont {Martinek}},
  \bibinfo {author} {\bibfnamefont {J.~E.}\ \bibnamefont {Grose}}, \bibinfo
  {author} {\bibfnamefont {L.~A.~K.}\ \bibnamefont {Donev}}, \bibinfo {author}
  {\bibfnamefont {P.~L.}\ \bibnamefont {McEuen}}, \ and\ \bibinfo {author}
  {\bibfnamefont {D.~C.}\ \bibnamefont {Ralph}},\ }\href@noop {} {\bibfield
  {journal} {\bibinfo  {journal} {Science}\ }\textbf {\bibinfo {volume}
  {306}},\ \bibinfo {pages} {86} (\bibinfo {year} {2004})}\BibitemShut
  {NoStop}%
\bibitem [{\citenamefont {Yu}\ \emph {et~al.}(2005)\citenamefont {Yu},
  \citenamefont {Keane}, \citenamefont {Ciszek}, \citenamefont {Cheng},
  \citenamefont {Tour}, \citenamefont {Baruah}, \citenamefont {Pederson},\ and\
  \citenamefont {Natelson}}]{Natelson2005}%
  \BibitemOpen
  \bibfield  {author} {\bibinfo {author} {\bibfnamefont {L.~H.}\ \bibnamefont
  {Yu}}, \bibinfo {author} {\bibfnamefont {Z.~K.}\ \bibnamefont {Keane}},
  \bibinfo {author} {\bibfnamefont {J.~W.}\ \bibnamefont {Ciszek}}, \bibinfo
  {author} {\bibfnamefont {L.}~\bibnamefont {Cheng}}, \bibinfo {author}
  {\bibfnamefont {J.~M.}\ \bibnamefont {Tour}}, \bibinfo {author}
  {\bibfnamefont {T.}~\bibnamefont {Baruah}}, \bibinfo {author} {\bibfnamefont
  {M.~R.}\ \bibnamefont {Pederson}}, \ and\ \bibinfo {author} {\bibfnamefont
  {D.}~\bibnamefont {Natelson}},\ }\href@noop {} {\bibfield  {journal}
  {\bibinfo  {journal} {Phys. Rev. Lett.}\ }\textbf {\bibinfo {volume} {95}},\
  \bibinfo {pages} {256803} (\bibinfo {year} {2005})}\BibitemShut {NoStop}%
\bibitem [{\citenamefont {Osorio}\ \emph {et~al.}(2007)\citenamefont {Osorio},
  \citenamefont {{O'Neill}}, \citenamefont {Wegewijs}, \citenamefont
  {{Stuhr-Hansen}}, \citenamefont {Paaske}, \citenamefont {Bj\o{}rnholm},\ and\
  \citenamefont {{van der Zant}}}]{Osorio2007}%
  \BibitemOpen
  \bibfield  {author} {\bibinfo {author} {\bibfnamefont {E.~A.}\ \bibnamefont
  {Osorio}}, \bibinfo {author} {\bibfnamefont {K.}~\bibnamefont {{O'Neill}}},
  \bibinfo {author} {\bibfnamefont {M.}~\bibnamefont {Wegewijs}}, \bibinfo
  {author} {\bibfnamefont {N.}~\bibnamefont {{Stuhr-Hansen}}}, \bibinfo
  {author} {\bibfnamefont {J.}~\bibnamefont {Paaske}}, \bibinfo {author}
  {\bibfnamefont {T.}~\bibnamefont {Bj\o{}rnholm}}, \ and\ \bibinfo {author}
  {\bibfnamefont {H.~S.~J.}\ \bibnamefont {{van der Zant}}},\ }\href@noop {}
  {\bibfield  {journal} {\bibinfo  {journal} {Nano Lett.}\ }\textbf {\bibinfo
  {volume} {7}},\ \bibinfo {pages} {3336} (\bibinfo {year} {2007})}\BibitemShut
  {NoStop}%
\bibitem [{\citenamefont {Mugarza}\ \emph {et~al.}(2011)\citenamefont
  {Mugarza}, \citenamefont {Krull}, \citenamefont {Robles}, \citenamefont
  {Stepanow}, \citenamefont {Ceballos},\ and\ \citenamefont
  {Gambardella}}]{Mugarza2011}%
  \BibitemOpen
  \bibfield  {author} {\bibinfo {author} {\bibfnamefont {A.}~\bibnamefont
  {Mugarza}}, \bibinfo {author} {\bibfnamefont {C.}~\bibnamefont {Krull}},
  \bibinfo {author} {\bibfnamefont {R.}~\bibnamefont {Robles}}, \bibinfo
  {author} {\bibfnamefont {S.}~\bibnamefont {Stepanow}}, \bibinfo {author}
  {\bibfnamefont {G.}~\bibnamefont {Ceballos}}, \ and\ \bibinfo {author}
  {\bibfnamefont {P.}~\bibnamefont {Gambardella}},\ }\href@noop {} {\bibfield
  {journal} {\bibinfo  {journal} {Nature Communications}\ }\textbf {\bibinfo
  {volume} {8}},\ \bibinfo {pages} {490} (\bibinfo {year} {2011})}\BibitemShut
  {NoStop}%
\bibitem [{\citenamefont {Rakhmilevitch}\ \emph {et~al.}(2014)\citenamefont
  {Rakhmilevitch}, \citenamefont {Korytar}, \citenamefont {Bagrets},
  \citenamefont {Evers},\ and\ \citenamefont {Tal}}]{Rakhmilevitch2013}%
  \BibitemOpen
  \bibfield  {author} {\bibinfo {author} {\bibfnamefont {D.}~\bibnamefont
  {Rakhmilevitch}}, \bibinfo {author} {\bibfnamefont {R.}~\bibnamefont
  {Korytar}}, \bibinfo {author} {\bibfnamefont {A.}~\bibnamefont {Bagrets}},
  \bibinfo {author} {\bibfnamefont {F.}~\bibnamefont {Evers}}, \ and\ \bibinfo
  {author} {\bibfnamefont {O.}~\bibnamefont {Tal}},\ }\href@noop {} {\bibfield
  {journal} {\bibinfo  {journal} {Phys. Rev. Lett.}\ }\textbf {\bibinfo
  {volume} {113}},\ \bibinfo {pages} {236603} (\bibinfo {year}
  {2014})}\BibitemShut {NoStop}%
\bibitem [{\citenamefont {Wagner}\ \emph {et~al.}(2013)\citenamefont {Wagner},
  \citenamefont {Kisslinger}, \citenamefont {Ballmann}, \citenamefont
  {Schramm}, \citenamefont {Chandrasekar}, \citenamefont {Bodenstein},
  \citenamefont {Fuhr}, \citenamefont {Secker}, \citenamefont {Fink},
  \citenamefont {Ruben},\ and\ \citenamefont {Weber}}]{Wagner2013}%
  \BibitemOpen
  \bibfield  {author} {\bibinfo {author} {\bibfnamefont {S.}~\bibnamefont
  {Wagner}}, \bibinfo {author} {\bibfnamefont {F.}~\bibnamefont {Kisslinger}},
  \bibinfo {author} {\bibfnamefont {S.}~\bibnamefont {Ballmann}}, \bibinfo
  {author} {\bibfnamefont {F.}~\bibnamefont {Schramm}}, \bibinfo {author}
  {\bibfnamefont {R.}~\bibnamefont {Chandrasekar}}, \bibinfo {author}
  {\bibfnamefont {R.}~\bibnamefont {Bodenstein}}, \bibinfo {author}
  {\bibfnamefont {O.}~\bibnamefont {Fuhr}}, \bibinfo {author} {\bibfnamefont
  {D.}~\bibnamefont {Secker}}, \bibinfo {author} {\bibfnamefont
  {K.}~\bibnamefont {Fink}}, \bibinfo {author} {\bibfnamefont {M.}~\bibnamefont
  {Ruben}}, \ and\ \bibinfo {author} {\bibfnamefont {H.~B.}\ \bibnamefont
  {Weber}},\ }\href@noop {} {\bibfield  {journal} {\bibinfo  {journal} {Nat.
  Nanotechnology}\ }\textbf {\bibinfo {volume} {8}},\ \bibinfo {pages} {575}
  (\bibinfo {year} {2013})}\BibitemShut {NoStop}%
\bibitem [{\citenamefont {{Goldhaber-Gordon}}\ \emph
  {et~al.}(1998{\natexlab{a}})\citenamefont {{Goldhaber-Gordon}}, \citenamefont
  {Shtrikman}, \citenamefont {Mahalu}, \citenamefont {{Abusch-Magder}},
  \citenamefont {Meirav},\ and\ \citenamefont {Kastner}}]{Goldhaber1998}%
  \BibitemOpen
  \bibfield  {author} {\bibinfo {author} {\bibfnamefont {D.}~\bibnamefont
  {{Goldhaber-Gordon}}}, \bibinfo {author} {\bibfnamefont {H.}~\bibnamefont
  {Shtrikman}}, \bibinfo {author} {\bibfnamefont {D.}~\bibnamefont {Mahalu}},
  \bibinfo {author} {\bibfnamefont {D.}~\bibnamefont {{Abusch-Magder}}},
  \bibinfo {author} {\bibfnamefont {U.}~\bibnamefont {Meirav}}, \ and\ \bibinfo
  {author} {\bibfnamefont {M.~A.}\ \bibnamefont {Kastner}},\ }\href@noop {}
  {\bibfield  {journal} {\bibinfo  {journal} {Nature}\ }\textbf {\bibinfo
  {volume} {391}},\ \bibinfo {pages} {156} (\bibinfo {year}
  {1998}{\natexlab{a}})}\BibitemShut {NoStop}%
\bibitem [{\citenamefont {Cronenwett}\ \emph {et~al.}(1998)\citenamefont
  {Cronenwett}, \citenamefont {Oosterkamp},\ and\ \citenamefont
  {Kouwenhoven}}]{Cronenwett1998}%
  \BibitemOpen
  \bibfield  {author} {\bibinfo {author} {\bibfnamefont {S.~M.}\ \bibnamefont
  {Cronenwett}}, \bibinfo {author} {\bibfnamefont {T.~H.}\ \bibnamefont
  {Oosterkamp}}, \ and\ \bibinfo {author} {\bibfnamefont {L.~P.}\ \bibnamefont
  {Kouwenhoven}},\ }\href@noop {} {\bibfield  {journal} {\bibinfo  {journal}
  {Science}\ }\textbf {\bibinfo {volume} {281}},\ \bibinfo {pages} {540}
  (\bibinfo {year} {1998})}\BibitemShut {NoStop}%
\bibitem [{\citenamefont {van~der Wiel}\ \emph {et~al.}(2000)\citenamefont
  {van~der Wiel}, \citenamefont {De~Franceschi}, \citenamefont {Elzerman},
  \citenamefont {Fujisawa}, \citenamefont {Tarucha},\ and\ \citenamefont
  {Kouwenhoven}}]{Wiel2000}%
  \BibitemOpen
  \bibfield  {author} {\bibinfo {author} {\bibfnamefont {W.~G.}\ \bibnamefont
  {van~der Wiel}}, \bibinfo {author} {\bibfnamefont {S.}~\bibnamefont
  {De~Franceschi}}, \bibinfo {author} {\bibfnamefont {J.~M.}\ \bibnamefont
  {Elzerman}}, \bibinfo {author} {\bibfnamefont {T.}~\bibnamefont {Fujisawa}},
  \bibinfo {author} {\bibfnamefont {S.}~\bibnamefont {Tarucha}}, \ and\
  \bibinfo {author} {\bibfnamefont {L.~P.}\ \bibnamefont {Kouwenhoven}},\
  }\href@noop {} {\bibfield  {journal} {\bibinfo  {journal} {Science}\ }\textbf
  {\bibinfo {volume} {282}},\ \bibinfo {pages} {2105} (\bibinfo {year}
  {2000})}\BibitemShut {NoStop}%
\bibitem [{\citenamefont {Jeong}\ \emph {et~al.}(2001)\citenamefont {Jeong},
  \citenamefont {Chang},\ and\ \citenamefont {Melloch}}]{Jeong2001}%
  \BibitemOpen
  \bibfield  {author} {\bibinfo {author} {\bibfnamefont {H.}~\bibnamefont
  {Jeong}}, \bibinfo {author} {\bibfnamefont {A.~M.}\ \bibnamefont {Chang}}, \
  and\ \bibinfo {author} {\bibfnamefont {M.~R.}\ \bibnamefont {Melloch}},\
  }\href@noop {} {\bibfield  {journal} {\bibinfo  {journal} {Science}\ }\textbf
  {\bibinfo {volume} {293}},\ \bibinfo {pages} {2221} (\bibinfo {year}
  {2001})}\BibitemShut {NoStop}%
\bibitem [{\citenamefont {H\"artle}\ \emph {et~al.}(2013)\citenamefont
  {H\"artle}, \citenamefont {Cohen}, \citenamefont {Reichman},\ and\
  \citenamefont {Millis}}]{Hartle2013b}%
  \BibitemOpen
  \bibfield  {author} {\bibinfo {author} {\bibfnamefont {R.}~\bibnamefont
  {H\"artle}}, \bibinfo {author} {\bibfnamefont {G.}~\bibnamefont {Cohen}},
  \bibinfo {author} {\bibfnamefont {D.~R.}\ \bibnamefont {Reichman}}, \ and\
  \bibinfo {author} {\bibfnamefont {A.~J.}\ \bibnamefont {Millis}},\
  }\href@noop {} {\bibfield  {journal} {\bibinfo  {journal} {Phys. Rev. B}\
  }\textbf {\bibinfo {volume} {88}},\ \bibinfo {pages} {235426} (\bibinfo
  {year} {2013})}\BibitemShut {NoStop}%
\bibitem [{\citenamefont {Hell}\ \emph {et~al.}(2015)\citenamefont {Hell},
  \citenamefont {Sothmann}, \citenamefont {Leijnse}, \citenamefont {Wegewijs},\
  and\ \citenamefont {K\"onig}}]{Hell2014}%
  \BibitemOpen
  \bibfield  {author} {\bibinfo {author} {\bibfnamefont {M.}~\bibnamefont
  {Hell}}, \bibinfo {author} {\bibfnamefont {B.}~\bibnamefont {Sothmann}},
  \bibinfo {author} {\bibfnamefont {M.}~\bibnamefont {Leijnse}}, \bibinfo
  {author} {\bibfnamefont {M.~R.}\ \bibnamefont {Wegewijs}}, \ and\ \bibinfo
  {author} {\bibfnamefont {J.}~\bibnamefont {K\"onig}},\ }\href@noop {}
  {\bibfield  {journal} {\bibinfo  {journal} {Phys. Rev. B}\ }\textbf {\bibinfo
  {volume} {91}},\ \bibinfo {pages} {195404} (\bibinfo {year}
  {2015})}\BibitemShut {NoStop}%
\bibitem [{\citenamefont {Jin}\ \emph {et~al.}(2008)\citenamefont {Jin},
  \citenamefont {Zheng},\ and\ \citenamefont {Yan}}]{Jin2008}%
  \BibitemOpen
  \bibfield  {author} {\bibinfo {author} {\bibfnamefont {J.}~\bibnamefont
  {Jin}}, \bibinfo {author} {\bibfnamefont {X.}~\bibnamefont {Zheng}}, \ and\
  \bibinfo {author} {\bibfnamefont {Y.}~\bibnamefont {Yan}},\ }\href@noop {}
  {\bibfield  {journal} {\bibinfo  {journal} {J. Chem. Phys.}\ }\textbf
  {\bibinfo {volume} {128}},\ \bibinfo {pages} {234703} (\bibinfo {year}
  {2008})}\BibitemShut {NoStop}%
\bibitem [{\citenamefont {van~der Wiel}\ \emph {et~al.}(2002)\citenamefont
  {van~der Wiel}, \citenamefont {De~Franceschi}, \citenamefont {Elzerman},
  \citenamefont {Fujisawa}, \citenamefont {Tarucha},\ and\ \citenamefont
  {Kouwenhoven}}]{Wiel2002}%
  \BibitemOpen
  \bibfield  {author} {\bibinfo {author} {\bibfnamefont {W.~G.}\ \bibnamefont
  {van~der Wiel}}, \bibinfo {author} {\bibfnamefont {S.}~\bibnamefont
  {De~Franceschi}}, \bibinfo {author} {\bibfnamefont {J.~M.}\ \bibnamefont
  {Elzerman}}, \bibinfo {author} {\bibfnamefont {T.}~\bibnamefont {Fujisawa}},
  \bibinfo {author} {\bibfnamefont {S.}~\bibnamefont {Tarucha}}, \ and\
  \bibinfo {author} {\bibfnamefont {L.~P.}\ \bibnamefont {Kouwenhoven}},\
  }\href@noop {} {\bibfield  {journal} {\bibinfo  {journal} {Rev. Mod. Phys.}\
  }\textbf {\bibinfo {volume} {75}},\ \bibinfo {pages} {1} (\bibinfo {year}
  {2002})}\BibitemShut {NoStop}%
\bibitem [{\citenamefont {Keller}\ \emph {et~al.}(2014)\citenamefont {Keller},
  \citenamefont {Amasha}, \citenamefont {Weymann}, \citenamefont {Moca},
  \citenamefont {Rau}, \citenamefont {Katine}, \citenamefont {Shtrikman},
  \citenamefont {Zarand},\ and\ \citenamefont
  {{Goldhaber-Gordon}}}]{Keller2014}%
  \BibitemOpen
  \bibfield  {author} {\bibinfo {author} {\bibfnamefont {A.~J.}\ \bibnamefont
  {Keller}}, \bibinfo {author} {\bibfnamefont {S.}~\bibnamefont {Amasha}},
  \bibinfo {author} {\bibfnamefont {I.}~\bibnamefont {Weymann}}, \bibinfo
  {author} {\bibfnamefont {C.~P.}\ \bibnamefont {Moca}}, \bibinfo {author}
  {\bibfnamefont {I.~G.}\ \bibnamefont {Rau}}, \bibinfo {author} {\bibfnamefont
  {J.~A.}\ \bibnamefont {Katine}}, \bibinfo {author} {\bibfnamefont
  {H.}~\bibnamefont {Shtrikman}}, \bibinfo {author} {\bibfnamefont
  {G.}~\bibnamefont {Zarand}}, \ and\ \bibinfo {author} {\bibfnamefont
  {D.}~\bibnamefont {{Goldhaber-Gordon}}},\ }\href@noop {} {\bibfield
  {journal} {\bibinfo  {journal} {Nat. Phys.}\ }\textbf {\bibinfo {volume}
  {10}},\ \bibinfo {pages} {145} (\bibinfo {year} {2014})}\BibitemShut
  {NoStop}%
\bibitem [{Note1()}]{Note1}%
  \BibitemOpen
  \bibinfo {note} {Using units where the elementary charge $e=1$, $\hbar =1$
  and the Boltzmann constant $k_\protect \text {B}=1$.}\BibitemShut {Stop}%
\bibitem [{\citenamefont {H\"artle}\ and\ \citenamefont
  {Millis}(2014)}]{Hartle2014}%
  \BibitemOpen
  \bibfield  {author} {\bibinfo {author} {\bibfnamefont {R.}~\bibnamefont
  {H\"artle}}\ and\ \bibinfo {author} {\bibfnamefont {A.~J.}\ \bibnamefont
  {Millis}},\ }\href@noop {} {\bibfield  {journal} {\bibinfo  {journal} {Phys.
  Rev. B}\ }\textbf {\bibinfo {volume} {90}},\ \bibinfo {pages} {245426}
  (\bibinfo {year} {2014})}\BibitemShut {NoStop}%
\bibitem [{\citenamefont {Kashcheyevs}\ \emph {et~al.}(2007)\citenamefont
  {Kashcheyevs}, \citenamefont {Schiller}, \citenamefont {Aharony},\ and\
  \citenamefont {Entin-Wohlman}}]{Kashcheyevs2007}%
  \BibitemOpen
  \bibfield  {author} {\bibinfo {author} {\bibfnamefont {V.}~\bibnamefont
  {Kashcheyevs}}, \bibinfo {author} {\bibfnamefont {A.}~\bibnamefont
  {Schiller}}, \bibinfo {author} {\bibfnamefont {A.}~\bibnamefont {Aharony}}, \
  and\ \bibinfo {author} {\bibfnamefont {O.}~\bibnamefont {Entin-Wohlman}},\
  }\href@noop {} {\bibfield  {journal} {\bibinfo  {journal} {Phys. Rev. B}\
  }\textbf {\bibinfo {volume} {75}},\ \bibinfo {pages} {115313} (\bibinfo
  {year} {2007})}\BibitemShut {NoStop}%
\bibitem [{\citenamefont {H\"artle}\ \emph {et~al.}(2015)\citenamefont
  {H\"artle}, \citenamefont {Cohen}, \citenamefont {Reichman},\ and\
  \citenamefont {Millis}}]{Hartle2015}%
  \BibitemOpen
  \bibfield  {author} {\bibinfo {author} {\bibfnamefont {R.}~\bibnamefont
  {H\"artle}}, \bibinfo {author} {\bibfnamefont {G.}~\bibnamefont {Cohen}},
  \bibinfo {author} {\bibfnamefont {D.~R.}\ \bibnamefont {Reichman}}, \ and\
  \bibinfo {author} {\bibfnamefont {A.~J.}\ \bibnamefont {Millis}},\
  }\href@noop {} {\bibfield  {journal} {\bibinfo  {journal} {Phys. Rev. B}\
  }\textbf {\bibinfo {volume} {92}},\ \bibinfo {pages} {085430} (\bibinfo
  {year} {2015})}\BibitemShut {NoStop}%
\bibitem [{\citenamefont {May}(2002)}]{May02}%
  \BibitemOpen
  \bibfield  {author} {\bibinfo {author} {\bibfnamefont {V.}~\bibnamefont
  {May}},\ }\href@noop {} {\bibfield  {journal} {\bibinfo  {journal} {Phys.
  Rev. B}\ }\textbf {\bibinfo {volume} {66}},\ \bibinfo {pages} {245411}
  (\bibinfo {year} {2002})}\BibitemShut {NoStop}%
\bibitem [{\citenamefont {Mitra}\ \emph {et~al.}(2004)\citenamefont {Mitra},
  \citenamefont {Aleiner},\ and\ \citenamefont {Millis}}]{Mitra04}%
  \BibitemOpen
  \bibfield  {author} {\bibinfo {author} {\bibfnamefont {A.}~\bibnamefont
  {Mitra}}, \bibinfo {author} {\bibfnamefont {I.}~\bibnamefont {Aleiner}}, \
  and\ \bibinfo {author} {\bibfnamefont {A.~J.}\ \bibnamefont {Millis}},\
  }\href@noop {} {\bibfield  {journal} {\bibinfo  {journal} {Phys. Rev. B}\
  }\textbf {\bibinfo {volume} {69}},\ \bibinfo {pages} {245302} (\bibinfo
  {year} {2004})}\BibitemShut {NoStop}%
\bibitem [{\citenamefont {Lehmann}\ \emph {et~al.}(2004)\citenamefont
  {Lehmann}, \citenamefont {Kohler}, \citenamefont {May},\ and\ \citenamefont
  {{H\"anggi}}}]{Lehmann04}%
  \BibitemOpen
  \bibfield  {author} {\bibinfo {author} {\bibfnamefont {J.}~\bibnamefont
  {Lehmann}}, \bibinfo {author} {\bibfnamefont {S.}~\bibnamefont {Kohler}},
  \bibinfo {author} {\bibfnamefont {V.}~\bibnamefont {May}}, \ and\ \bibinfo
  {author} {\bibfnamefont {P.}~\bibnamefont {{H\"anggi}}},\ }\href@noop {}
  {\bibfield  {journal} {\bibinfo  {journal} {J. Chem. Phys.}\ }\textbf
  {\bibinfo {volume} {121}},\ \bibinfo {pages} {2278} (\bibinfo {year}
  {2004})}\BibitemShut {NoStop}%
\bibitem [{\citenamefont {Harbola}\ \emph {et~al.}(2006)\citenamefont
  {Harbola}, \citenamefont {Esposito},\ and\ \citenamefont
  {Mukamel}}]{Harbola2006}%
  \BibitemOpen
  \bibfield  {author} {\bibinfo {author} {\bibfnamefont {U.}~\bibnamefont
  {Harbola}}, \bibinfo {author} {\bibfnamefont {M.}~\bibnamefont {Esposito}}, \
  and\ \bibinfo {author} {\bibfnamefont {S.}~\bibnamefont {Mukamel}},\
  }\href@noop {} {\bibfield  {journal} {\bibinfo  {journal} {Phys. Rev. B}\
  }\textbf {\bibinfo {volume} {74}},\ \bibinfo {pages} {235309} (\bibinfo
  {year} {2006})}\BibitemShut {NoStop}%
\bibitem [{\citenamefont {Koch}\ \emph {et~al.}(2006)\citenamefont {Koch},
  \citenamefont {Semmelhack}, \citenamefont {{von Oppen}},\ and\ \citenamefont
  {Nitzan}}]{Semmelhack}%
  \BibitemOpen
  \bibfield  {author} {\bibinfo {author} {\bibfnamefont {J.}~\bibnamefont
  {Koch}}, \bibinfo {author} {\bibfnamefont {M.}~\bibnamefont {Semmelhack}},
  \bibinfo {author} {\bibfnamefont {F.}~\bibnamefont {{von Oppen}}}, \ and\
  \bibinfo {author} {\bibfnamefont {A.}~\bibnamefont {Nitzan}},\ }\href@noop {}
  {\bibfield  {journal} {\bibinfo  {journal} {Phys. Rev. B}\ }\textbf {\bibinfo
  {volume} {73}},\ \bibinfo {pages} {155306} (\bibinfo {year}
  {2006})}\BibitemShut {NoStop}%
\bibitem [{\citenamefont {Siddiqui}\ \emph {et~al.}(2007)\citenamefont
  {Siddiqui}, \citenamefont {Ghosh},\ and\ \citenamefont {Datta}}]{Siddiqui}%
  \BibitemOpen
  \bibfield  {author} {\bibinfo {author} {\bibfnamefont {L.}~\bibnamefont
  {Siddiqui}}, \bibinfo {author} {\bibfnamefont {A.~W.}\ \bibnamefont {Ghosh}},
  \ and\ \bibinfo {author} {\bibfnamefont {S.}~\bibnamefont {Datta}},\
  }\href@noop {} {\bibfield  {journal} {\bibinfo  {journal} {Phys. Rev. B}\
  }\textbf {\bibinfo {volume} {76}},\ \bibinfo {pages} {085433} (\bibinfo
  {year} {2007})}\BibitemShut {NoStop}%
\bibitem [{\citenamefont {Timm}(2008)}]{Timm08}%
  \BibitemOpen
  \bibfield  {author} {\bibinfo {author} {\bibfnamefont {C.}~\bibnamefont
  {Timm}},\ }\href@noop {} {\bibfield  {journal} {\bibinfo  {journal} {Phys.
  Rev. B}\ }\textbf {\bibinfo {volume} {77}},\ \bibinfo {pages} {195416}
  (\bibinfo {year} {2008})}\BibitemShut {NoStop}%
\bibitem [{\citenamefont {H\"artle}\ \emph {et~al.}(2009)\citenamefont
  {H\"artle}, \citenamefont {Benesch},\ and\ \citenamefont {Thoss}}]{Hartle09}%
  \BibitemOpen
  \bibfield  {author} {\bibinfo {author} {\bibfnamefont {R.}~\bibnamefont
  {H\"artle}}, \bibinfo {author} {\bibfnamefont {C.}~\bibnamefont {Benesch}}, \
  and\ \bibinfo {author} {\bibfnamefont {M.}~\bibnamefont {Thoss}},\
  }\href@noop {} {\bibfield  {journal} {\bibinfo  {journal} {Phys. Rev. Lett.}\
  }\textbf {\bibinfo {volume} {102}},\ \bibinfo {pages} {146801} (\bibinfo
  {year} {2009})}\BibitemShut {NoStop}%
\bibitem [{\citenamefont {H\"artle}\ and\ \citenamefont
  {Thoss}(2011)}]{Hartle2010b}%
  \BibitemOpen
  \bibfield  {author} {\bibinfo {author} {\bibfnamefont {R.}~\bibnamefont
  {H\"artle}}\ and\ \bibinfo {author} {\bibfnamefont {M.}~\bibnamefont
  {Thoss}},\ }\href@noop {} {\bibfield  {journal} {\bibinfo  {journal} {Phys.
  Rev. B}\ }\textbf {\bibinfo {volume} {83}},\ \bibinfo {pages} {115414}
  (\bibinfo {year} {2011})}\BibitemShut {NoStop}%
\bibitem [{\citenamefont {K\"onig}\ and\ \citenamefont
  {Martinek}(2003)}]{Martinek2003}%
  \BibitemOpen
  \bibfield  {author} {\bibinfo {author} {\bibfnamefont {J.}~\bibnamefont
  {K\"onig}}\ and\ \bibinfo {author} {\bibfnamefont {J.}~\bibnamefont
  {Martinek}},\ }\href@noop {} {\bibfield  {journal} {\bibinfo  {journal}
  {Phys. Rev. Lett.}\ }\textbf {\bibinfo {volume} {90}},\ \bibinfo {pages}
  {166602} (\bibinfo {year} {2003})}\BibitemShut {NoStop}%
\bibitem [{\citenamefont {Wunsch}\ \emph {et~al.}(2005)\citenamefont {Wunsch},
  \citenamefont {Braun}, \citenamefont {K\"onig},\ and\ \citenamefont
  {Pfannkuche}}]{Wunsch2005}%
  \BibitemOpen
  \bibfield  {author} {\bibinfo {author} {\bibfnamefont {B.}~\bibnamefont
  {Wunsch}}, \bibinfo {author} {\bibfnamefont {M.}~\bibnamefont {Braun}},
  \bibinfo {author} {\bibfnamefont {J.}~\bibnamefont {K\"onig}}, \ and\
  \bibinfo {author} {\bibfnamefont {D.}~\bibnamefont {Pfannkuche}},\
  }\href@noop {} {\bibfield  {journal} {\bibinfo  {journal} {Phys. Rev. B}\
  }\textbf {\bibinfo {volume} {72}},\ \bibinfo {pages} {205319} (\bibinfo
  {year} {2005})}\BibitemShut {NoStop}%
\bibitem [{\citenamefont {Pedersen}\ \emph {et~al.}(2007)\citenamefont
  {Pedersen}, \citenamefont {Lassen}, \citenamefont {Wacker},\ and\
  \citenamefont {Hettler}}]{Hettler2007}%
  \BibitemOpen
  \bibfield  {author} {\bibinfo {author} {\bibfnamefont {J.~N.}\ \bibnamefont
  {Pedersen}}, \bibinfo {author} {\bibfnamefont {B.}~\bibnamefont {Lassen}},
  \bibinfo {author} {\bibfnamefont {A.}~\bibnamefont {Wacker}}, \ and\ \bibinfo
  {author} {\bibfnamefont {M.~H.}\ \bibnamefont {Hettler}},\ }\href@noop {}
  {\bibfield  {journal} {\bibinfo  {journal} {Phys. Rev. B}\ }\textbf {\bibinfo
  {volume} {75}},\ \bibinfo {pages} {235314} (\bibinfo {year}
  {2007})}\BibitemShut {NoStop}%
\bibitem [{\citenamefont {Trocha}\ \emph {et~al.}(2009)\citenamefont {Trocha},
  \citenamefont {Weymann},\ and\ \citenamefont {Barnas}}]{Trocha2009}%
  \BibitemOpen
  \bibfield  {author} {\bibinfo {author} {\bibfnamefont {P.}~\bibnamefont
  {Trocha}}, \bibinfo {author} {\bibfnamefont {I.}~\bibnamefont {Weymann}}, \
  and\ \bibinfo {author} {\bibfnamefont {J.}~\bibnamefont {Barnas}},\
  }\href@noop {} {\bibfield  {journal} {\bibinfo  {journal} {Phys. Rev. B}\
  }\textbf {\bibinfo {volume} {80}},\ \bibinfo {pages} {165333} (\bibinfo
  {year} {2009})}\BibitemShut {NoStop}%
\bibitem [{\citenamefont {Donarini}\ \emph {et~al.}(2010)\citenamefont
  {Donarini}, \citenamefont {Begemann},\ and\ \citenamefont
  {Grifoni}}]{Donarini2010}%
  \BibitemOpen
  \bibfield  {author} {\bibinfo {author} {\bibfnamefont {A.}~\bibnamefont
  {Donarini}}, \bibinfo {author} {\bibfnamefont {G.}~\bibnamefont {Begemann}},
  \ and\ \bibinfo {author} {\bibfnamefont {M.}~\bibnamefont {Grifoni}},\
  }\href@noop {} {\bibfield  {journal} {\bibinfo  {journal} {Phys. Rev. B}\
  }\textbf {\bibinfo {volume} {82}},\ \bibinfo {pages} {125451} (\bibinfo
  {year} {2010})}\BibitemShut {NoStop}%
\bibitem [{\citenamefont {Holm}\ \emph {et~al.}(2008)\citenamefont {Holm},
  \citenamefont {Jorgensen}, \citenamefont {{Grove-Rasmussen}}, \citenamefont
  {Paaske}, \citenamefont {Flensberg},\ and\ \citenamefont
  {Lindelof}}]{Paaske2008}%
  \BibitemOpen
  \bibfield  {author} {\bibinfo {author} {\bibfnamefont {J.~V.}\ \bibnamefont
  {Holm}}, \bibinfo {author} {\bibfnamefont {H.~I.}\ \bibnamefont {Jorgensen}},
  \bibinfo {author} {\bibfnamefont {K.}~\bibnamefont {{Grove-Rasmussen}}},
  \bibinfo {author} {\bibfnamefont {J.}~\bibnamefont {Paaske}}, \bibinfo
  {author} {\bibfnamefont {K.}~\bibnamefont {Flensberg}}, \ and\ \bibinfo
  {author} {\bibfnamefont {P.~E.}\ \bibnamefont {Lindelof}},\ }\href@noop {}
  {\bibfield  {journal} {\bibinfo  {journal} {Phys. Rev. B}\ }\textbf {\bibinfo
  {volume} {77}},\ \bibinfo {pages} {161406} (\bibinfo {year}
  {2008})}\BibitemShut {NoStop}%
\bibitem [{\citenamefont {Begemann}\ \emph {et~al.}(2010)\citenamefont
  {Begemann}, \citenamefont {Koller}, \citenamefont {Grifoni},\ and\
  \citenamefont {Paaske}}]{Begemann2010}%
  \BibitemOpen
  \bibfield  {author} {\bibinfo {author} {\bibfnamefont {G.}~\bibnamefont
  {Begemann}}, \bibinfo {author} {\bibfnamefont {S.}~\bibnamefont {Koller}},
  \bibinfo {author} {\bibfnamefont {M.}~\bibnamefont {Grifoni}}, \ and\
  \bibinfo {author} {\bibfnamefont {J.}~\bibnamefont {Paaske}},\ }\href@noop {}
  {\bibfield  {journal} {\bibinfo  {journal} {Phys. Rev. B}\ }\textbf {\bibinfo
  {volume} {82}},\ \bibinfo {pages} {045316} (\bibinfo {year}
  {2010})}\BibitemShut {NoStop}%
\bibitem [{\citenamefont {{Goldhaber-Gordon}}\ \emph
  {et~al.}(1998{\natexlab{b}})\citenamefont {{Goldhaber-Gordon}}, \citenamefont
  {G\"ores}, \citenamefont {Kastner}, \citenamefont {Shtrikman}, \citenamefont
  {Mahalu},\ and\ \citenamefont {Meirav}}]{Goldhaber1998b}%
  \BibitemOpen
  \bibfield  {author} {\bibinfo {author} {\bibfnamefont {D.}~\bibnamefont
  {{Goldhaber-Gordon}}}, \bibinfo {author} {\bibfnamefont {J.}~\bibnamefont
  {G\"ores}}, \bibinfo {author} {\bibfnamefont {M.~A.}\ \bibnamefont
  {Kastner}}, \bibinfo {author} {\bibfnamefont {H.}~\bibnamefont {Shtrikman}},
  \bibinfo {author} {\bibfnamefont {D.}~\bibnamefont {Mahalu}}, \ and\ \bibinfo
  {author} {\bibfnamefont {U.}~\bibnamefont {Meirav}},\ }\href@noop {}
  {\bibfield  {journal} {\bibinfo  {journal} {Phys. Rev. Lett.}\ }\textbf
  {\bibinfo {volume} {81}},\ \bibinfo {pages} {5225} (\bibinfo {year}
  {1998}{\natexlab{b}})}\BibitemShut {NoStop}%
\end{thebibliography}
\end{document}